\documentclass[sts]{imsart}

\RequirePackage{amsthm,amsmath,amsfonts,amssymb}
\RequirePackage[numbers]{natbib}

\usepackage{threeparttable}
\usepackage{epsfig, verbatim,enumerate, wrapfig}
\usepackage{graphicx}
\usepackage{epstopdf}
\usepackage{subfig}
\usepackage{mathrsfs, amsmath}
\usepackage{tgtermes}  	% like times
\usepackage{scalerel}

\usepackage{amsmath,amssymb, amsfonts,euscript,mathrsfs, mathtools}
\usepackage{algorithm,algpseudocode}

\renewcommand{\thealgorithm}{}

\usepackage{pdfpages}

\usepackage{amsthm}

\usepackage{todonotes}
\usepackage{mdwlist}

\usepackage{enumitem}

\usepackage[left=1in, right=1in,top=1in, bottom=1in]{geometry}
\usepackage[colorlinks,bookmarksopen,bookmarksnumbered,citecolor=blue,urlcolor=blue]{hyperref} 
\usepackage[utf8x]{inputenc}
\usepackage{float}
\usepackage{multirow}
\floatstyle{plaintop}
\usepackage{amsmath,amssymb, amsfonts,euscript,mathrsfs, mathtools}

\usepackage{bbm}

\usepackage{adjustbox} %%rotate tables

\usepackage{pdfpages}
\usepackage{epstopdf}

\usepackage{tikz-cd}

\newtheorem{theorem}{Theorem}

\allowdisplaybreaks

\startlocaldefs
\endlocaldefs

\begin{document}

\begin{frontmatter}
\title{Inference for BART with Multinomial Outcomes}
%\title{A sample article title with some additional note\thanksref{T1}}
%\runtitle{???}
%\thankstext{T1}{A sample of additional note to the title.}

\begin{aug}
\author[A]{\fnms{Yizhen} \snm{Xu}\ead[label=e1]{yxu143@jhu.edu}},
\author[B]{\fnms{Joseph W.} \snm{Hogan}\ead[label=e2]{jwh@brown.edu}}
\author[C]{\fnms{Michael J.} \snm{Daniels}\ead[label=e3]{mdaniels@stat.ufl.edu}}
\author[D]{\fnms{Rami} \snm{Kantor}\ead[label=e4]{rkantor@brown.edu}}
\and
\author[E]{\fnms{Kantor} \snm{Ann Mwangi}\ead[label=e5]{annwsum@gmail.com}}

%%%%%%%%%%%%%%%%%%%%%%%%%%%%%%%%%%%%%%%%%%%%%%
%% Addresses                                %%
%%%%%%%%%%%%%%%%%%%%%%%%%%%%%%%%%%%%%%%%%%%%%%
\address[A]{Department of Biostatistics, Johns Hopkins University, 615 N Wolfe St, Baltimore MD, U.S.A., \printead{e1}.}
\address[B]{Department of Biostatistics, Brown University,121 S. Main Street, Providence RI, U.S.A., \printead{e2}.}
\address[C]{Department of Statistics, University of Florida, Gainesville FL, U.S.A, \printead{e3}.}
\address[D]{Division of Infectious Diseases, Brown University, Providence RI, U.S.A, \printead{e4}.}
\address[E]{Academic Model Providing Access to Healthcare (AMPATH), Eldoret, Kenya, \printead{e5}.}
\address[E]{College of Health Sciences, School of Medicine, Moi University, Eldoret, Kenya, \printead{e5}.}
\end{aug}

\begin{abstract}
The multinomial probit Bayesian additive regression trees (MPBART) framework was proposed by \citep{kindo_multinomial_2016} (KD), approximating the latent utilities in the multinomial probit (MNP) model with BART \citep{chipman_bart:_2010}. Compared to multinomial logistic models, MNP does not assume independent alternatives and the correlation structure among alternatives can be specified through multivariate Gaussian distributed latent utilities. We introduce two new algorithms for fitting the MPBART and show that the theoretical mixing rates of our proposals are equal or superior to the existing algorithm in KD. Through simulations, we explore the robustness of the methods to the choice of reference level, imbalance in outcome frequencies, and the specifications of prior hyperparameters for the utility error term. The work is motivated by the application of generating posterior predictive distributions for mortality and engagement in care among HIV-positive patients based on electronic health records (EHRs) from the Academic Model Providing Access to Healthcare (AMPATH) in Kenya. In both the application and simulations, we observe better performance using our proposals as compared to KD in terms of MCMC convergence rate and posterior predictive accuracy. The package for implementation is publicly available at \url{https://github.com/yizhenxu/GcompBART}.
\end{abstract}


\end{frontmatter}
%%%%%%%%%%%%%%%%%%%%%%%%%%%%%%%%%%%%%%%%%%%%%%
%%%% Main text entry area:

\section{Introduction}\label{sec:1} 

Bayesian additive regression trees (BART)  \citep{chipman_bart:_2010} is a flexible nonparametric Bayesian approach for regression on a recursively binary-partitioned predictor space; it uses sum-of-trees to model the mean function such that nonlinearities and interactions along with additive effects are naturally accounted for, and regularization priors are imposed to favor shallow trees to reduce over-fitting. There has been considerable literature on extending BART to various types of problems given BART's predictive power for continuous and binary outcomes \citep{sparapani_nonparametric_2016,waldmann_genome-wide_2016,low-kam_bayesian_2015}. We specifically consider the extension of BART to multinomial probit models \citep{imai_bayesian_2005} (MNP). Existing BART-related work has developed efficient Markov chain Monte Carlo (MCMC) algorithms for Gaussian likelihoods, which naturally adapt to frameworks with Gaussian-distributed latent variables. However, careful consideration of data augmentation (DA) schemes is needed for the computational efficiency of implementing BART for categorical outcomes. The main contributions of this paper are to provide a detailed review of sampling algorithms for parameter expansion that are based on DA schemes and to introduce a set of new MCMC algorithms for multinomial probit BART (MPBART).

Our work is motivated by the predictive modeling of the progression of patients' retention and mortality through the HIV care cascade \citep{who_meeting_2012,gardner_spectrum_2011}, adopting the regression framework proposed by \citep{lee_state-space_2017} for modeling patients' engagement in care over time as transitions among categorical outcome states conditional on patients' clinical history. The HIV care cascade is a conceptual model that outlines essential stages of the HIV care continuum: (a) HIV diagnosis through testing, (b) linkage to care, (c) engagement in care, (d) initiation of antiviral therapy (ART) through retention, and (e) sustained suppression of viral load; it has been widely used as a monitoring and evaluation tool for improving and managing HIV health care systems. We will demonstrate and compare different algorithms of MPBART for this application in Section \ref{app2}.

MNP \citep{imai_bayesian_2005} and multinomial logistic \citep{mcfadden_conditional_1974} (MNL) models are widely used tools for predicting and describing the relationships of explanatory variables to categorical outcomes. \citep{kindo_multinomial_2016} proposed the MPBART framework that uses BART to fit models to the Gaussian latent variables in the MNP. Related work incorporating BART into categorical response models is introduced by \citep{murray_log-linear_2020}, where BART is extended to log-linear models that includes multinomial logistic BART (MLBART). Both MNP and MNL can be derived from a latent variable framework, where each outcome category is modeled by a utility which is a function of the covariates, and the outcome is the utility-maximizing category. MNP and MNL assume the utilities to be multivariate Gaussian distributed and independent extreme-value distributed, respectively. The MNP is appealing because the independence assumption under the MNL may be counterintuitive in practice \citep{steenburgh_invariant_2008}; the MNP does not have this restriction. This property is also present for MLBART and MPBART. We will show that allowing non-zero correlations between utilities is an important feature of the latent variable distribution and can have a substantial impact on predictive accuracy.

There are two difficulties in sampling from posterior distributions for MNP. First, a closed-form expression for the outcome's marginal probabilities is not available; second, identifiability of the MNP requires constraints on the covariance matrix of the utilities, hindering specification of conjugate distributions and making posterior sampling more difficult. There has been considerable work on Bayesian sampling techniques to address the computational issues of the MNP \citep{albert_bayesian_1993, mcculloch_exact_1994,mcculloch_bayesian_2000,nobile_hybrid_1998,imai_bayesian_2005} based on DA-related methods. The original DA algorihtm \citep{tanner_calculation_1987} is a stochastic generalization of the EM algorithm \citep{dempster_maximum_1977}. Marginal data augmentation (MDA) \citep{meng_seeking_1999,liu_parameter_1999, van_dyk_art_2001} generalizes and accelerates the DA algorithm via parameter expansion such that full conditionals are easier to sample from and expansion parameter(s) are subsequently marginalized over. Heuristically, the MDA Gibbs sampler can traverse the parameter space more efficiently with the extra variation induced by the expansion parameter(s), resulting in possible computational gains, including a faster mixing rate \citep{meng_seeking_1999, liu_parameter_1999}. \citep{li_using_2018} provided an example for posterior sampling of a correlation matrix via parameter expansion. For the MNP, sampling from the constrained model parameter space is difficult because the full conditionals do not have a simple closed form; the MDA scheme circumvents the difficulty and allows an easier and more efficient joint sampling of expansion parameter and transformed model parameters.
\citep{imai_bayesian_2005} unified several previous proposals under the umbrella of MDA, examined different prior specifications for model parameters, and outlined two adaptations of the MDA scheme for posterior sampling of the MNP based on parameter expansion.

Building upon the work of \citep{imai_bayesian_2005}, \citep{kindo_multinomial_2016} proposed an algorithm, which we refer to as KD, for fitting the MPBART. However, the existing package of KD has been shown to have problems \citep{murray_log-linear_2020}; our implementation of KD shows further drawbacks such as oversized posterior trees from potential overfitting, difficulty in posterior convergence, and lack of robustness to the choice of reference levels. We propose two alternative procedures for fitting the MPBART that have simpler algorithmic structure, improved convergence in the sum-of-trees and the covariance matrix, and a better mixing rate when the Markov chain reaches equilibrium. In particular, our algorithms show better out-of-sample accuracy and stability in predictive tasks under various settings when evaluated in terms of posterior means and posterior modes; the posterior mode accuracy is used as the comparison metric in \citep{kindo_multinomial_2016}. The intuition behind our proposals is to fit the sum-of-trees in a constrained parameter space to reduce disruptions to the stochastic search of posterior trees, resulting in faster convergence of the Markov chain.

In every step of the Gibbs sampler, the MDA scheme requires (1) the joint sampling of expansion parameter(s) and transformed model parameters, and (2) the marginalization over the expansion parameter. However, the two actions are not always feasible for complicated Gibbs sampling problems. For example, sampling the functional mean component jointly with an expansion parameter in an MPBART algorithm is difficult because posterior trees are sampled by stochastic search. Algorithms for the MNP and MPBART generally fall under the category of the partially marginalized augmentation (PMA) samplers \citep{van_dyk_marginal_2010}, which relaxes the fully marginalized structure of the MDA and expects an improvement in convergence rate when more steps involve joint sampling and marginalizing components of the expansion parameter(s). We will show that KD and one of our proposals are essentially the MPBART generalizations of the two procedures proposed in \citep{imai_bayesian_2005}; the main difference between the two MPBART algorithms is that the latter does not employ augmentation in the sampling of model coefficients for the mean component of random utilities. In the context of MPBART, this is contrary to the intuition about PMA samplers and shows that: when Metropolis-Hastings or stochastic search is involved in complicated samplers, particular steps may be sensitive to the incorporation of expansion parameter(s); as such, more considerations in algorithm design are needed.

This paper is structured as follows. Section \ref{notation} introduces the formulation of MNP and MPBART frameworks; Section \ref{DA} reviews sampling schemes for the MNP, including DA and MDA;  Section \ref{algs} introduces the existing and proposed algorithms for fitting the MPBART; and Section \ref{compare} theoretically compares different MPBART algorithms in terms of the mixing rate under stationarity. Section \ref{sim2} provides empirical investigations into the MPBART algorithms on simulated data under different settings. Section \ref{app2} conducts further comparisons of the algorithms on a real-world dataset. Section \ref{last2} summarizes the conclusions.

\section{Method}\label{med2}

\subsection{General Background}\label{notation}

For the categorical outcome $S$, which takes value in $\{0,\ldots,C\}$, the general latent variable framework for multinomial models assumes that $S$ is a manifestation of unobserved latent utilities $Z=(Z_0,\ldots,Z_C)\in \mathbb{R}^{C+1}$ , where $S(Z) = \text{argmax}_l Z_l,$
i.e.\ $S=k$ if $Z_k\ge Z_l$ for all $l\neq k$. In general, $C$ is the number of outcome levels minus one. The framework requires normalization for identifiability because $S$ is invariant to a translation or a scaling (by a positive constant) of $Z$. Without loss of generality, we assume that the reference outcome category is 0; the normalization is achieved by modeling $S$ as a function of latent variables $W = (W_{1},\ldots,W_{C})\in \mathbb{R}^{C}$, such that $W_l = Z_l-Z_0$ and 
\begin{align}
S(W) &= \left\{\begin{matrix}
l & \text{if max} (W)=W_{l} \ge 0\\ 
0 & \text{if max} (W) < 0.
\end{matrix}\right. \label{latent}
\end{align}
The MNP models $W$ in terms of covariates $X$ and accounts for correlation across outcome levels by assuming $W$ follows a multivariate normal model,
\begin{align}
W(X) \sim MVN(G(X;\theta) , \Sigma ),\label{MVN}
\end{align}
where  \[G(X;\theta) = (G_1(X;\theta_1),\ldots,G_{C}(X;\theta_{C})), \theta = (\theta_1,\ldots,\theta_C),\] and $\Sigma = \{\sigma_{ij}\}$ is a $C\times C$ positive definite symmetric matrix. 

Identifiability of the model requires normalizing the scale of $W$ because by definition the outcome $S$ is invariant to a multiplication of $W$ by any positive constant. From \eqref{MVN}, the normalization for scale occurs by imposing a constraint on the covariance matrix $\Sigma$, such as $\text{trace} (\Sigma) = C$  \citep{burgette_trace_2012}. To illustrate, suppose there are latent variables $\widetilde{W}$ of the following form,
\begin{align}
\widetilde{W}(X)  \sim MVN(G(X;\widetilde{\theta}),\widetilde{\Sigma}),
\label{unidentify}
\end{align}
where $\widetilde{W}(X) = \alpha W(X)$, $G(X;\widetilde{\theta}) = \alpha G(X;\theta)$, $\widetilde{\Sigma} = \alpha^2 \Sigma$, and $\alpha>0$. By \eqref{latent}, $\widetilde{W}$ and $W$ yield the same $S$. However, if $\Sigma$ satisfies the trace constraint, $W$ is the normalized counterpart of $\widetilde{W}$ and $\alpha^2=\text{trace}(\widetilde{\Sigma})/C$ is a positive scalar that ensures a one-to-one mapping from $W$ to $\widetilde{W}$.

Direct posterior sampling of \eqref{MVN} is difficult due to the constraint on $\Sigma$. A technique for easier sampling is to augment the parameter space such that it is possible to specify a conjugate prior and target parameters can be obtained by converting samples back to the normalized scale. The obvious choice of augmented parameter space is the one without the normalization for scale, i.e.\ $(\widetilde{W}, \widetilde{\theta}, \widetilde{\Sigma})$ in \eqref{unidentify}. \citep{imai_bayesian_2005} suggest a constrained inverse Wishart prior for $\Sigma$ such that its joint distribution with $\alpha^2$ is equivalent to the unconstrained covariance matrix having prior distribution $\widetilde{\Sigma} \sim \text{inv-Wishart}(\nu, \Psi)$. This makes it possible to sample easily from the conditional posterior of $\widetilde{\Sigma}$. Setting $\nu=C+1$ and $\Psi$ to be an identity matrix is equivalent to sampling the corresponding correlations of $\widetilde{\Sigma}$ from a uniform distribution. When $\nu > C+1$, the expectation of $\widetilde{\Sigma}$ has a closed form $E(\widetilde{\Sigma}) = \frac{\Psi}{\nu-C-1}$.    

The classic framework of MNP assumes a linear model specification for each $W_l(X)$, i.e.\ $G_l(X;\theta_l) =  X\theta_l$ for $l=1,\ldots, C$. \citep{kindo_multinomial_2016} proposed MPBART to increase the predictive power and the flexibility in dealing with complicated nonlinear and interaction effects. The innovative idea is to approximate each mean component of $W(X)$ using a sum of $m$ trees, $G_l(X;\theta_l)=\sum_{k=1}^m g(X; \theta_{lk}),$
where $l=1,\ldots,C$ and $\theta_{lk}$ is the set of parameters corresponding to the $k$th binary tree for the $l$th latent variable, $W_l(X)$. MPBART uses the same Bayesian regularization prior on the trees to restrict over-fitting as in \citep{chipman_bart:_2010}. An important contribution of \citep{kindo_multinomial_2016} is deriving from \eqref{MVN} the conditional distribution for Gibbs sampling of each individual tree, and embedding it into the backfitting procedure of BART. See \citep{chipman_bayesian_1998,chipman_bart:_2010} for details on the BART backfitting procedure.

\subsection{Review of Data Augmentation} \label{DA}

Suppose $y$ and $\phi$ represent the augmented data and model parameters, respectively. The goal of data augmentation (DA) schemes is to draw samples from $(y,\phi)$. The sampling algorithm proposed by \citep{kindo_multinomial_2016} for MPBART heavily relies on Imai \& van Dyk's \citep{imai_bayesian_2005} work on fitting the MNP, which explores different Gibbs samplers of $(W, \theta, \Sigma)$ under the umbrella of marginal data augmentation (MDA) \citep{meng_seeking_1999,liu_parameter_1999}, an extension and improvement of the DA algorithm \citep{tanner_calculation_1987}. This section provides a brief overview of relevant developments on the DA algorithm for fitting the MPBART in Section \ref{algs}. 

{\em Basic data augmentation.}
To begin with, we illustrate the simple task of sampling $(y,\phi)$ under the DA algorithm of \citep{tanner_calculation_1987} :\\

\noindent \textbf{Scheme [DA]}
\begin{enumerate}[leftmargin=*,itemsep=0mm]
\item Draw $y\sim f(y|\phi)$.
\item Draw $\phi\sim f(\phi|y)$.
\end{enumerate}

{\em Marginalized data augmentation (MDA).}
The basic idea of MDA versus DA is to expand the model and overparameterize $f(y,\phi)$ to $f(y,\phi,\alpha)$; the expansion parameter $\alpha$ often corresponds to a transformation of $y$ and/or $\phi$. For example, $\alpha$ may index a transformation of $y$ to $\widetilde{y} = t_\alpha(y)$ where $t_\alpha$ is one-to-one and differentiable, expanding the model from $f(y,\phi)$ to $f(\widetilde{y}, \phi,\alpha)$. The choice to sample from $f(y,\phi,\alpha)$ or $f(\widetilde{y},\phi,\alpha)$ depends on the specific model, and they are usually interchangeable. This approach is appealing when sampling from $f(y,\alpha|\phi)$ or $f(\tilde{y},\alpha|\phi)$ is easier than the sampling of $y$ alone. \citep{liu_parameter_1999} and \citep{meng_seeking_1999} simultaneously developed MDA. \citep{liu_parameter_1999} provided theoretical results on the convergence rate of the MDA. \citep{meng_seeking_1999} introduced the MDA under two augmentation schemes, \textit{grouping} and \textit{collapsing} \citep{liu_collapsed_1994,liu_covariance_1994}; both procedures lead to the same distribution of $(y,\phi)$ as Scheme [DA]. 

{\em MDA with grouping.}
The grouping scheme samples conditionally on the expansion parameter $\alpha$, while the collapsing scheme integrates $\alpha$ out from the joint distribution. MDA under the grouping scheme is preferred when the sampling of $y$ or $\phi$ jointly with $\alpha$ is easier than that in Scheme [DA]. For example, when $f(\phi | y, \alpha)$ is easier to sample than $f(\phi | y)$, and $f(y,\alpha | \phi)$ is easy to sample, the sampler can ``group'' $y$ and $\alpha$ together and treats them as a single component,\\

\noindent \textbf{Scheme [MDA-G]}
\begin{enumerate}[leftmargin=*,itemsep=0mm]
\item Draw $(y,\alpha)\sim f(y,\alpha| \phi)$.
\item Draw  $\phi\sim f(\phi |y,\alpha)$.
\end{enumerate}

{\em MDA with collapsing.}
MDA under the collapsing scheme ``collapses down'' $\alpha$ by integrating it out from the joint distributions, i.e.\ $y\sim f(y|\phi) = \int f(y|\phi, \alpha)f(\alpha | \phi) d\alpha$ and $\phi\sim f(\phi|y) = \int f(\phi |y,\alpha)f(\alpha|y)d\alpha$. The implementation is as follows:\\

\noindent \textbf{Scheme [MDA-C]}
\begin{enumerate}[leftmargin=*,itemsep=0mm]
\item Draw $(y,\alpha)\sim f(y,\alpha| \phi)$ by $\alpha\sim f(\alpha|\phi)$ and $y\sim f(y|\phi,\alpha)$.
\item Draw $(\phi,\alpha)\sim f(\phi, \alpha|y)$ by $\alpha\sim f(\alpha|y)$ and $\phi\sim f(\phi|y,\alpha)$.
\end{enumerate}

Notice that the newly sampled $\alpha$ is discarded in each step of the Scheme [MDA-C]. In practice, it may be reasonable to assume a priori independence between $\phi$ and $\alpha$ because $\phi$ are parameters identified from the observed data, which does not contain information on $\alpha$. Furthermore, given that transforming the augmented data $y$ is of interest, it may be true that the conditional sampling of model parameters $\phi$ is more plausible under $\widetilde{y}$ than $y$. Accordingly, Scheme [MDA-C] can be rewritten as:\\

\noindent \textbf{Scheme [MDA-C']}
\begin{enumerate}[leftmargin=*,itemsep=0mm]
\item Draw $(\widetilde{y},\alpha)$ by drawing $\alpha\sim f(\alpha)$ and then $y\sim f(y|\phi,\alpha)$, and compute $\widetilde{y} = t_\alpha(y)$.
\item Draw $(\phi,\alpha)$ by drawing $\alpha\sim f(\alpha|\widetilde{y})$ and then $\phi\sim f(\phi|\widetilde{y},\alpha)$. 
\end{enumerate}

The $f(\alpha)$ and $f(\alpha|\widetilde{y})$ are the prior and posterior (under the transformed augmented data) of $\alpha$, respectively. The optimality of MDA under the collapsing scheme (Scheme [MDA-C]) over the DA algorithm (Scheme [DA]) in terms of convergence rate is proven in \citep{meng_seeking_1999} and \citep{liu_parameter_1999}. \citep{liu_parameter_1999} also introduced Scheme [MDA-LW], which is equivalent to Scheme [MDA-C'] in terms of the sampling distribution and rate of convergence. This scheme is implicitly applied in the algorithms for fitting the MNP and MPBART, typically in the normalization of model parameters after each round of Gibbs sampling. Structurally, Scheme [MDA-LW] is in the form of Scheme [DA] with an additional intermediate step, which makes more clear the connection between the MDA and the DA algorithm:\\

\noindent \textbf{Scheme [MDA-LW]}
\begin{enumerate}[leftmargin=*,itemsep=0mm]
\item Draw $y \sim f(y|\phi)$.
\item Draw $\alpha_1 \sim f(\alpha)$, compute $\widetilde{y} = t_{\alpha_1}(y)$; draw $\alpha_2 \sim f(\alpha|\widetilde{y})$, compute $y'=t^{-1}_{\alpha_2}(\widetilde{y})$.
\item Draw $\phi\sim f(\phi|y').$
\end{enumerate}

Note that $y$ and $y'$ follow the same distribution. The intuition behind the improvement of Scheme [MDA-LW] compared to the DA algorithm is that the intermediate step of sampling from $y'$ allows the sampler for $\phi$ to explore the expanded model space with more freedom.

\subsection{Data Augmentation for the MNP} \label{DA-MNP}

For fitting the MNP, \citep{imai_bayesian_2005} introduced two algorithms for the Gibbs sampling of $(W,\theta,\Sigma)$, which we refer as IvD1 and IvD2. The IvD1 modifies Scheme [MDA-C'] by expanding the model to $(\widetilde{W},\widetilde{\theta},\widetilde{\Sigma},\alpha)$ such that $\widetilde{W}$ and $(\widetilde{\theta},\widetilde{\Sigma})$ correspond to $\widetilde{y}$ and $\phi$, respectively, and $\alpha = (\alpha_1, \alpha_2, \alpha_3)$:\\

\noindent \textbf{Scheme [IvD1]}
\begin{enumerate}[leftmargin=*,itemsep=0mm]
\item Draw $(\widetilde{W},\alpha_1)$ by drawing $\alpha_1\sim f(\alpha|\Sigma)$ and then $W\sim f(W|\theta,\Sigma)$, and compute $\widetilde{W} = \alpha_1 W$.
\item Draw $(\widetilde{\theta},\alpha_2)$ by drawing $\alpha_2\sim f(\alpha|\widetilde{W},\Sigma)$ and then $\widetilde{\theta}\sim f(\widetilde{\theta}|\alpha_2,\widetilde{W},\Sigma)$, and compute $\theta = \widetilde{\theta}/\alpha_2$.
\item Draw $(\widetilde{\Sigma},\alpha_3)$ by $\widetilde{\Sigma} \sim f(\widetilde{\Sigma}  | \widetilde{W}  - X\widetilde{\theta})$ and compute $\alpha_3=\sqrt{\text{trace}(\widetilde{\Sigma})/C}.$
\end{enumerate}

Using $\widetilde{\Sigma}$ and $\alpha_3$ from Step 3, we can compute the normalized covariance matrix $\Sigma = \widetilde{\Sigma}/\alpha^2_3$ and use it in Steps 1 and 2 of the next round of posterior sampling; this is analogous to having $\alpha_3$ index a one-to-one mapping from the expanded model space ($\widetilde{\Sigma}$) to the normalized space ($\Sigma$). Steps 1 and 3 in Scheme [IvD1] collapse down $\alpha_1$ and $\alpha_3$, but Scheme [IvD1] is not a direct implementation of the MDA as in Scheme [MDA-C'] because Step 1 is conditional on $\theta$, or equivalently $(\widetilde{\theta},\alpha_2)$ where $\theta = \widetilde{\theta}/\alpha_2$. Hence, Step 2 does not integrate out (collapse down) $\alpha_2$. 

Standard MDA  (Schemes [MDA-C] and [MDA-C']) are collapsed Gibbs samplers that integrate out expansion parameter(s) by redrawing and discarding $\alpha$ in every step. Scheme [IvD1] is in fact a partially marginalized augmentation (PMA) \citep{van_dyk_marginal_2010} procedure that relaxes the restrictive structure of full marginalization in MDA. PMA allows the conditional distribution in a $k$th step of the Gibbs sampler to depend on expansion parameter(s) drawn in other steps. Algorithms for fitting the MPBART in Section \ref{algs}  are generally also PMA procedures.

IvD1 can also be viewed from a different perspective. Due to the linearity in model specification of the MNP, i.e.\ $G_l(X;\theta_l) = X\theta_l$ for $l=1,\ldots,C$, the linear relationship between $\theta$ and $\widetilde{\theta}$ holds in Step 2 of Scheme [IvD1], and it is equivalent to direct sampling of $\theta$ from $f(\theta|\widetilde{W}/\alpha_2, \Sigma)$. Hence, IvD1 can be rearranged as follows \\

\noindent \textbf{Scheme [IvD1']}
\begin{enumerate}[leftmargin=*,itemsep=0mm]
\item Draw $W\sim f(W|\theta,\Sigma)$.
\item Draw $\alpha_1\sim f(\alpha|\Sigma)$, compute $\widetilde{W} = \alpha_1 W$; draw $\alpha_2\sim f(\alpha|\widetilde{W},\Sigma)$, compute $W'=\widetilde{W}/\alpha_2$.
\item Draw $\theta \sim f(\theta|W', \Sigma)$.
\item Draw $\Sigma$ by $\widetilde{\Sigma} \sim f(\widetilde{\Sigma}  | \widetilde{W}  - X\widetilde{\theta})$, compute \\
$\alpha_3=\sqrt{\text{trace}(\widetilde{\Sigma})/C}$, and $\Sigma = \widetilde{\Sigma}/\alpha^2_3$, where $\widetilde{\theta}=\alpha_2\theta$.
\end{enumerate}

The first three steps are equivalent to sampling $f(W,\theta|\Sigma)$ in Scheme [MDA-LW]. Step 4 collapses down $\alpha_3$, but the fact that Step 4 is conditional on $(\alpha_1,\alpha_2)$ through $(\widetilde{W}, \widetilde{\theta})$ makes IvD1 not a collapsed Gibbs sampler collectively. IvD2 is given as follows:\\

\noindent \textbf{Scheme [IvD2]}
\begin{enumerate}[leftmargin=*,itemsep=0mm]
\item Draw $(\widetilde{\epsilon},\alpha_1)$ by $\alpha_1\sim f(\alpha|\Sigma)$ and $W\sim f(W|\theta,\Sigma)$, compute $\widetilde{\epsilon} = \alpha_1 [W-G(X;\theta)]$.
\item Draw $(\Sigma,\alpha_3)$ by $\widetilde{\Sigma} \sim f(\widetilde{\Sigma}|\widetilde{\epsilon})$, compute \\
$\alpha_3=\sqrt{\text{trace}(\widetilde{\Sigma})/C}$, and $\Sigma = \widetilde{\Sigma}/\alpha^2_3$.
\item Draw $\theta\sim f(\theta|W,\Sigma)$.	
\end{enumerate}

IvD2 separates the sampling into two parts, $(\widetilde{\epsilon},\Sigma)$ and $\theta$; the first part utilizes the MDA under Scheme [MDA-C] and the second part is a standard Gibbs sampling draw. Theoretically, as stated in \citep{imai_bayesian_2005}, IvD1 and IvD2 have the same lag-one autocorrelation when the MCMC chain is stationary. However, they showed through numerical experiments that IvD1 is better than IvD2 in estimating the MNP in terms of being less sensitive to the starting values of $(\theta,\Sigma)$.

In the next section, we describe Kindo et al.'s\citep{kindo_multinomial_2016} algorithm (KD) and our two new proposals, and connect them to the schemes reviewed here.

\subsection{Algorithms for Posterior Sampling Algorithms of MPBART}\label{algs}

For ease of notation, let $W_{i,-j} = (W_{i1},\ldots,W_{i,j-1},$ $W_{i,j+1},\ldots,W_{iC})$ and let $\mu = G(X; \theta)$ be the sum-of-trees component under the normalization of scale. Kindo et al.'s algorithm for fitting the MPBART can be summarized as the following augmented Gibbs sampler:\\

\noindent \textbf{Algorithm [KD]}
\begin{enumerate}[leftmargin=*]
	
	\item Sample $\widetilde{W},\alpha^2_1|\mu,\Sigma, S$. 
	\begin{enumerate}[leftmargin=*,itemsep=0mm,label=(\alph*)]
		\item Draw $\alpha^2_1$ from its conditional prior $f(\alpha^2|\Sigma)= \text{trace}[\Psi \Sigma^{-1}]/\chi^2_{\nu C}$;
		\item for each $j$, update $W_{ij}$ conditional on $W_{i,-j}$, $\mu$, $\Sigma$, and the observed outcome $S_i$, from a truncated normal distribution; and
		\item  transform $W_i$ and $\Sigma$ to $\widetilde{W}_i = \alpha_1 W_i$ and $\widetilde{\Sigma}^*=\alpha^2_1\Sigma$.
	\end{enumerate}
	
	\item Sample $\widetilde{\theta}|\widetilde{W},\alpha^2_1,\Sigma$. 
	\begin{enumerate}[leftmargin=*,itemsep=0mm,label=(\alph*)]
		\item Draw $\widetilde{\theta}\sim f(\widetilde{\theta}|\widetilde{W},\widetilde{\Sigma}^*)$; and
		\item set  $ \widetilde{\mu} = G(X;\widetilde{\theta})$ and $ \mu = \widetilde{\mu}/\alpha_1$.
	\end{enumerate}
	
	\item Sample $\Sigma,\alpha^2_3 |\widetilde{W}, \widetilde{\theta}$. 
	\begin{enumerate}[leftmargin=*,itemsep=0mm,label=(\alph*)]
		\item Draw $\widetilde{\Sigma}\sim \text{Inv-Wishart}(N+\nu, \Psi+\sum^N_{i=1}\widetilde{\epsilon}_i \widetilde{\epsilon}_i^T)$,
		where $\widetilde{\epsilon}_i = \widetilde{W}_i-\widetilde{\mu}_i$; 
		\item set $\alpha^2_3=\text{trace}(\widetilde{\Sigma}) / C$; and
		\item set $\Sigma = \widetilde{\Sigma}/\alpha^2_3$ and $W = \mu + \frac{\widetilde{\epsilon}}{\alpha_3}$.
	\end{enumerate}

\end{enumerate}

Step 1 jointly samples from  $f(\widetilde{W},\alpha^2_1|\mu, \Sigma, S)$ by first drawing the expansion parameter $\alpha^2_1$ from its prior distribution $f(\alpha^2|\Sigma)$, and then computing $\widetilde{W}=\alpha^2 W$ where $W$ is sampled from $f(W|\mu, \Sigma, S)$. Step 1(a) samples $\alpha^2_1$ such that $\alpha^2_1/\text{trace}[\Psi \Sigma^{-1}]$ follows an inverse-chi-squared distribution with $\nu C$ degrees of freedom. Step 1(b) samples each $W_{ij}$ from a truncated normal distribution described in Appendix \ref{A11} based on \eqref{latent}, as the observed outcome $S_i$ imposes an interval constraint on $W_i$, e.g.\  if $S_i$ equals the reference level 0, then $W_{ij}$'s are right truncated at 0. Step 2 samples posterior trees across multivariate mean components by Gibbs sampling and each posterior tree is sampled as in regular BART. Step 3 computes $\alpha_3$ using the sampled $\widetilde{\Sigma}$ and then normalizes the scale of the model by Step 3(c).  

Notice that the sampling of model parameters $\widetilde{\theta}$ is conditional on $(\widetilde{W},\widetilde{\Sigma}^*)$, which is equivalent to conditioning on $(\widetilde{W},\alpha^2_1,\Sigma)$ or $(W,\alpha^2_1,\Sigma)$; this observation is essential to the analysis of Algorithm [KD] in Section \ref{compare}. Algorithm [KD] is closely related to IvD1 (Scheme [IvD1]) but different in that it does not update the expansion parameter $\alpha_2$ as in Step (b) of IvD1. This is analogous to having $\alpha_2$ in IvD1 set to the sampled value of $\alpha_1$ from Step (a). The reason for this modification is that the posterior tree parameters in BART, denoted by $\theta$, are drawn via stochastic search; it would be extremely challenging to derive an analytical expression for $f(\alpha|\widetilde{W},\Sigma)$ from $\int f(\alpha,\theta|\widetilde{W},\Sigma)d\theta$ as in MNP, since the specification is no longer linear in $\theta$.

In the first step, $\widetilde{W}$ is a scaled version of $W$ through $\widetilde{W} = \alpha_1 W$. From \eqref{unidentify}, fitting the sum-of-trees component to $\widetilde{W}$ is analogous to sampling the parameters in an un-normalized space. Posterior tree sampling in BART makes a one-step update on each tree from its previous state, by one of the following four types of proposals: GROW, PRUNE, CHANGE, and SWAP. Stochastic search in a massive space of possible tree structures can be challenging when $\widetilde{W}$, the quantity to which the sum-of-trees is fitting, is unstable. Heuristically, we would expect fitting the sum-of-trees component to $W$, which is a normalized quantity, instead of $\widetilde{W}$ to be more stable, induce better posterior convergence, and improve the prediction accuracy. Given these considerations, we modify Algorithm [KD] and propose the following: \\

\noindent \textbf{Algorithm [P1]}
\begin{enumerate}[leftmargin=*]
	
	\item Sample $W,\alpha^2_1|\mu,\Sigma,S$. 
	\begin{enumerate}[leftmargin=*,itemsep=0mm,label=(\alph*)]
		\item Draw $\alpha^2_1$ from its conditional prior $f(\alpha^2|\Sigma)= \text{trace}[\Psi \Sigma^{-1}]/\chi^2_{\nu C}$; 
		\item for each $j$, update $W_{ij}$ conditional on $W_{i,-j}$, $\mu$, $\Sigma$, and observed outcome $S_i$, from a truncated normal distribution; and
		\item transform $W_i$ to $\widetilde{W}_i = \alpha_1 W_i$.
	\end{enumerate}
	
	\item Sample $\theta|W,\Sigma$. Draw $\theta\sim f(\theta|W,\Sigma)$ and then set  $\mu=G(X;\theta)$.

	\item Sample $\Sigma,\alpha^2_3 |W,\alpha_1,\theta $. 
	\begin{enumerate}[leftmargin=*,itemsep=0mm,label=(\alph*)]
		\item Draw $\widetilde{\Sigma}\sim \text{Inv-Wishart}(N+\nu, \Psi+\sum^N_{i=1}\widetilde{\epsilon}_i \widetilde{\epsilon}_i^T)$,
		where $\widetilde{\epsilon}_i = \alpha_1(W_i-\mu_i)$;
		\item set $\alpha^2_3$ to $\text{trace}(\widetilde{\Sigma}) / C$; and
		\item set $\Sigma = \widetilde{\Sigma}/\alpha^2_3$ and $W = \mu + \frac{\widetilde{\epsilon}}{\alpha_3}$.
	\end{enumerate}

\end{enumerate}
In the first proposal (Algorithm [P1]), the expansion parameters $(\alpha_1,\alpha_3)$ do not affect the sampling of the trees in Step 2. If the order of Step 2 and 3 are swapped, it becomes Scheme [IvD2] in Section \ref{DA}. Algorithms [KD] and [P1] are the MPBART analogues of IvD1 and IvD2 for the MNP. \citep{imai_bayesian_2005} expected IvD1 to outperform IvD2 for the MNP and demonstrated through simulations. While for MPBART, we find Algorithm [P1] to be equal or superior to Algorithm [KD] theoretically (Section \ref{compare}) and computationally (Sections \ref{sim2} and \ref{app2}).

As an alternative to Algorithm [P1], we introduce another proposal, Algorithm [P2], which ``abandons'' the MDA framework and adopts a Scheme [MDA-LW]-like strategy only for Step 3. If we fix $\alpha_1$ to be 1, both Algorithms [KD] and [P1] simplify to Algorithm [P2]. We show in Appendix \ref{2eq3} that Algorithms [P1] and [P2] draw $\Sigma$ from approximately the same sampling distribution under certain conditions. \\

\noindent \textbf{Algorithm [P2]}
\begin{enumerate}[leftmargin=*]
	
	\item Sample $W|\mu,\Sigma, S$. For each $j$, update $W_{ij}$ conditional on $W_{i,-j}$, $\mu$, $\Sigma$,$S_i$ from a truncated normal distribution.
	\item Sample $\theta|W,\Sigma$. Draw $\theta\sim f(\theta|W,\Sigma)$ and then set  $\mu=G(X;\theta)$.
	\item Sample $\Sigma, \alpha^2_3|W,\theta$. 
	\begin{enumerate}[leftmargin=*,itemsep=0mm,label=(\alph*)]
		\item Draw $\widetilde{\Sigma}\sim \text{Inv-Wishart}(N+\nu, \Psi+\sum^N_{i=1}\epsilon_i \epsilon^T_i)$, where $\epsilon_i = W_i-\mu_i$;
		\item set $\alpha^2_3$ to $\text{trace}(\widetilde{\Sigma}) / C$; and
		\item set $\Sigma = \widetilde{\Sigma}/\alpha^2_3$ and $W = \mu + \frac{\epsilon}{\alpha_3}$.	
	\end{enumerate}

\end{enumerate}
Appendix \ref{Paper3Alg} provides more details on the implementation of the algorithms. Software for fitting both the KD and our proposed algorithms is available on the first author's github page (\url{https://github.com/yizhenxu/GcompBART}).

\subsection{Comparisons of Algorithms for MPBART}\label{compare}

In what follows, we assume the Markov chain of $(W,\theta,\Sigma)$ has reached stationary. \citep{liu_fraction_1994a} introduced the usage of diagrams that show dependency structures between two consecutive iterations for analyzing Bayesian algorithms. We do this for Algorithms [KD], [P1], and [P2], and derive their mixing rate in terms of autocovariances. We restate the algorithms under the expanded model $(W,\mu,\Sigma,\alpha)$, where $W$ is the normalized latent variables with distribution $MVN(\mu,\Sigma)$ and $\alpha$ is the expansion parameter: 
\begin{align*}
&\text{Algorithm [KD]:\quad}\\
&(W,\alpha_1)|\mu,\Sigma \quad\Rightarrow \quad \mu|(W,\alpha_1),\Sigma  \quad \Rightarrow \quad (\Sigma,\alpha_3)|(W,\alpha_1),\mu,\\
&\text{Algorithm [P1]:\hspace{1.4em}}\\
&(W,\alpha_1)|\mu,\Sigma \quad \Rightarrow \quad 
\mu|W,\Sigma  \quad \Rightarrow \quad 
(\Sigma,\alpha_3)|(W,\alpha_1),\mu,\\
&\text{Algorithm [P2]:\hspace{1.4em}}\\
&W|\mu,\Sigma \quad \Rightarrow \quad 
\mu|W,\Sigma  \quad \Rightarrow \quad 
(\Sigma,\alpha_3)|W,\mu,
\end{align*}
where $\alpha$'s are indexed as in Scheme [IvD1].

We make a few observations about these three algorithms: (a) Algorithm [KD] groups $W$ and $\alpha_1$ together, as in Scheme [MDA-G]; (b) Algorithm [P1] is structurally equivalent to Scheme [IvD2]; and (c) the sampling of the normalized covariance matrix in all three algorithms integrates out $\alpha_3$ as in Scheme [MDA-C], i.e.\ $\Sigma\sim \int f(\Sigma,\alpha|W,\mu)d\alpha$ in Algorithm [P2], and $\Sigma\sim \int f(\Sigma,\alpha|W,\alpha_1,\mu)d\alpha$ in Algorithms [KD] and [P1]. Based on these observations, we prove the dependency structure as diagrams in Figure \ref{diagram}. 

A common measure for quantifying the mixing rate of a Markov chain is the lag-1 autocorrelation; lower autocorrelation indicates a better mixing rate. Using the dependency diagrams, we argue that Algorithm [P2] has the best mixing rate when the Markov chain is stationary. 

\begin{theorem}\label{Thm}
Assuming the chain of MPBART parameters $(W,\mu,\Sigma,\alpha)$ has reached equilibrium, 
\begin{enumerate}
	\item For $\mu$, Algorithms [P1] and [P2] have the same lag-1 autocorrelation, which is no larger than that from Algorithm [KD];
	\item For $\Sigma$, Algorithms [KD] and [P1] have the same lag-1 autocorrelation, which is no less than that from Algorithm [P2].
\end{enumerate}
\end{theorem}
Proof: Appendix \ref{ThmProof}.

\section{Simulation}\label{sim2}

This simulation study will compare the three algorithms in terms of MCMC convergence and prediction accuracy. We consider two different metrics of predictive accuracy which we define as follows. For subject $i=1,\ldots,N$ in a data for calculating predictive accuracy, given covariates $X_i$ and the posterior samples of model parameters $\{\theta^{(j)},\Sigma^{(j)}|j=1,\ldots,J\}$, the posterior predictive distribution for $S_i$ can be represented by its $J$ posterior predictions, $\{\hat{S}^{(1)}_{i},\ldots,\hat{S}^{(J)}_{i}\}$, where
\begin{align}
\hat{S}^{(j)}_{i} &= \left\{\begin{matrix}
l & \text{if max} (\hat{W}^{(j)}_{i})= \hat{W}^{(j)}_{il} \ge 0\\ 
C & \text{if max} (\hat{W}^{(j)}_{i}) < 0,
\end{matrix}\right. \label{pred}
\end{align}
$\hat{W}^{(j)}_{i} = (\hat{W}^{(j)}_{i1},\ldots,\hat{W}^{(j)}_{i,C})$ is the vector of latent variables,
$\hat{W}^{(j)}_{i} \sim MVN(G(X_i;\theta^{(j)}) , \Sigma^{(j)} )$, and \[G(X_i;\theta^{(j)}) = (G_1(X_i;\theta^{(j)}_1),\ldots,G_{C}(X_i;\theta^{(j)}_{C})).\] 
Recall that each mean component is parameterized as sum of trees,
$G_l(X_i;\theta^{(j)}_l)=\sum_{k=1}^m g(X_i; \theta^{(j)}_{lk})$, where $l=1,\ldots,C$.

We use posterior percent agreement and posterior mode to assess prediction accuracy. 
Posterior percent agreement is the percent concordance between the observed outcome $S_i$ and the posterior predictions of the outcome$\{\hat{S}^{(1)}_{i},\ldots,\hat{S}^{(J)}_{i}\}$, averaged over $N$ observations and $J$ posterior samples,
\begin{align}
& \frac{1}{N}\sum^N_{i=1}\bigg\{\frac{1}{J}\sum^J_{j=1} \mathbbm{1}\{\hat{S}^{(j)}_{i} = S_i\}\bigg\}.\label{accmean}
\end{align}
We also summarize agreement between the observed outcome and
the posterior mode via
\begin{align}
& \frac{1}{N} \sum^N_{i=1} \mathbbm{1}\{\text{mode}_j ( \hat{S}_{i}^{(j)} )= S_i\} \label{accmode}
\end{align}
where $\text{mode}_j(\hat{S}_{i}^{(j)})  = \text{argmax}_{l\in \{0,\ldots,C\}} \sum^J_{j=1}\mathbbm{1}\{ \hat{S}^{(j)}_{i} = l\}$ is the most frequently observed
outcome prediction among the posterior predictive draws
 $\hat{S}_i^{(j)}$  for individual $i$. 
The accuracy measures are different in that \eqref{accmode} ignores the infrequent categories in MCMC sampling while \eqref{accmean} uses all posterior predictions. 

Numerical experiments for simulations use 30,000 posterior draws after a burn-in of 50,000 for each model.
We used a sum of 100 trees for estimating the mean component of each latent variable. Following \citep{chipman_bart:_2010},
we set prior probabilities for the posterior tree search to be 0.25, 0.25, 0.4, and 0.1 for tree GROWTH, PRUNE, CHANGE, and SWAP, respectively. 
We also used the regularization priors for trees as recommended in \citep{chipman_bart:_2010}  Prior specification of the latent variable covariance matrix assumes the scale matrix $\Psi$ 
has diagonal elements equal to 1.

For each simulation, we create a training set $\mathcal{D}_1$ and test set $\mathcal{D}_2$, each of size 5000. Under different specifications of the reference level and prior on the covariance matrix, we use Algorithms [KD], [P1], and [P2] on $\mathcal{D}_1$. For each set of posterior samples for each algorithm, the corresponding out-of-sample performance is calculated by using \eqref{accmean} and \eqref{accmode} on $\mathcal{D}_2$. 

We use two specifications similar to those used by \citep{kang_demystifying_2007} 
for generating the outcome distribution in $\mathcal{D}_1$ and $\mathcal{D}_2$. We set $C=2$ and assume a set of covariates  $(U,V)$ where $U=(U_1,\ldots,U_5) \overset{\text{iid}}{\sim} \text{Uniform}(0,1)$ and $V \sim \text{Uniform}(0,2)$, and set
\begin{eqnarray*}
G_1 &=& 15 \sin(\pi U_1U_2) + (U_3-0.5)^2 -10U_4 -5U_5.
\end{eqnarray*}
For Setting 1, we set
\begin{eqnarray*}
G_2 &= &(U_3-0.5)^3 - 20U_4U_5 +4V,
\end{eqnarray*}
which yields a relatively balanced distribution for the multinomial categories in $S$.
For Setting 2, we set
\begin{eqnarray*}
G_2 &=& (U_3-0.5)^2 - U_4U_5 +4V,
\end{eqnarray*}
which yields highly imbalanced distribution across the categories similar to our
application.  The covariance matrix is given by 
$\Sigma = \left(\begin{matrix}
1& 0.5\\
0.5 & 1
\end{matrix}\right)$
for both Settings.

\subsection{Setting 1}\label{setting1}

Under Setting 1, the frequency of the outcome alternatives for the first simulated dataset was 
$(2227,1288,1485)$ and $(2225,1256,1519)$ for the training and test set, respectively. Table \ref{Table1} compares the out-of-sample accuracy of Algorithms [KD], [P1], and [P2] for Setting 1 with the reference level set to 1, 2, and 3, and the prior $\widetilde{\Sigma}\sim \text{Inv-Wishart}(3,I_2)$. Algorithms [P1] and [P2] have similar accuracy, and are always better than Algorithm [KD]. The accuracy values are very different between the metrics in \eqref{accmean} and \eqref{accmode}; compared to Algorithm [KD], the improvement in Algorithms [P1] and [P2] is much larger under the percent agreement metric \eqref{accmean}. In addition, Algorithm [KD] is sensitive to the choice of reference level; its posterior agreement accuracy under-performs under reference level 1, while the other two algorithms stay relatively stable. 

Because the covariance $\Sigma$ of the latent variables partially determines the outcome distribution and affects the predictive accuracy, we also investigate how the three algorithms behave in estimating $\Sigma$ when the reference level for MPBART is the same as that being used for the data generation, i.e.\ under reference level 3. We look at how different prior specifications on $\Sigma$ affect the posterior accuracy and actual estimation of $\Sigma$. In particular, we consider $\text{Inv-Wishart}(\nu,\Psi)$ prior for $\widetilde{\Sigma}$ with $\Psi_{11} = \Psi_{22} = 1$, comparing a uniform ($\nu=C+1, \Psi_{12}=0$), negatively tilted ($\nu=C+3, \Psi_{12}=-0.5$), and positively tilted ($\nu=C+3, \Psi_{12}=0.5$) prior for the un-normalized covariance $\sigma_{12}$. Table \ref{Table3} compares the out-of-sample accuracy and Table \ref{Table4} reports the posterior mean and 95 $\%$ credible intervals of the normalized covariance matrix entries $\sigma_{11}$ and $\sigma_{12}$. Note that we do not look at $\sigma_{22}$ because $\Sigma$ is normalized on its scale, so $\sigma_{22} = \text{trace}(\Sigma) - \sigma_{11}$. 

\begin{figure*}
			\centering
			\includegraphics[scale=0.6]{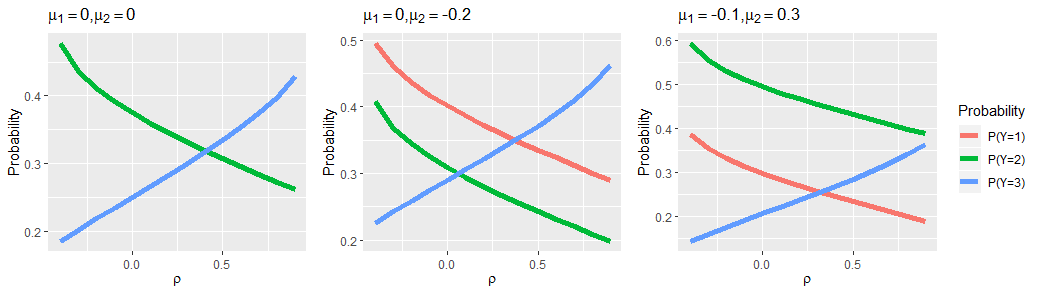}
			\caption{Outcome distribution as a function of $\rho$ under $(\mu_1,\mu_2)$ being (0,0), (0,-0.2), and (-0.1, 0.3). }
			\label{Sbysig12Plot}
\end{figure*}

Appendix \ref{Sbysig12} shows how $\sigma_{12}$ affects the outcome distribution, given $\sigma_{11} = \sigma_{22} = 1$. Conditional on $(G_1,G_2)$, $\sigma_{12}$ has a substantial effect on the outcome predictive distribution, as shown in Figure \ref{Sbysig12Plot}, and a negative $\sigma_{12}$ induces smaller reference level outcome probability $P(S = 3)$. On the other hand, having a negative estimated posterior mean of $\sigma_{12}$ as in Algorithm [KD] may lead to estimates of $G_1$ and $G_2$ that are systematically different from that in the true data generating mechanism, in which $\sigma_{12}$ is positive. 

It is clear from Table \ref{Table3} that compared to Algorithms [P1] and [P2], Algorithm [KD] is more sensitive to the prior specifications on $\Sigma$. Notably, Algorithm [KD] systematically gives a negative estimated posterior mean of the covariance $\sigma_{12}$, even though the empirical $\text{corr}(W_1,W_2)$ from the simulated training data is 0.424 and the true conditional correlation is $\text{corr}(W_1,W_2|G)$ is 0.5. On the other hand, the posterior estimate of $\text{cov}(W_1,W_2)$ is positive from Algorithms [P1] and [P2].

\begin{figure*}
	\centering
	\subfloat[$\nu=C+1$, $\Psi_{12}=0$, reference level 1]{\includegraphics[scale=0.17]{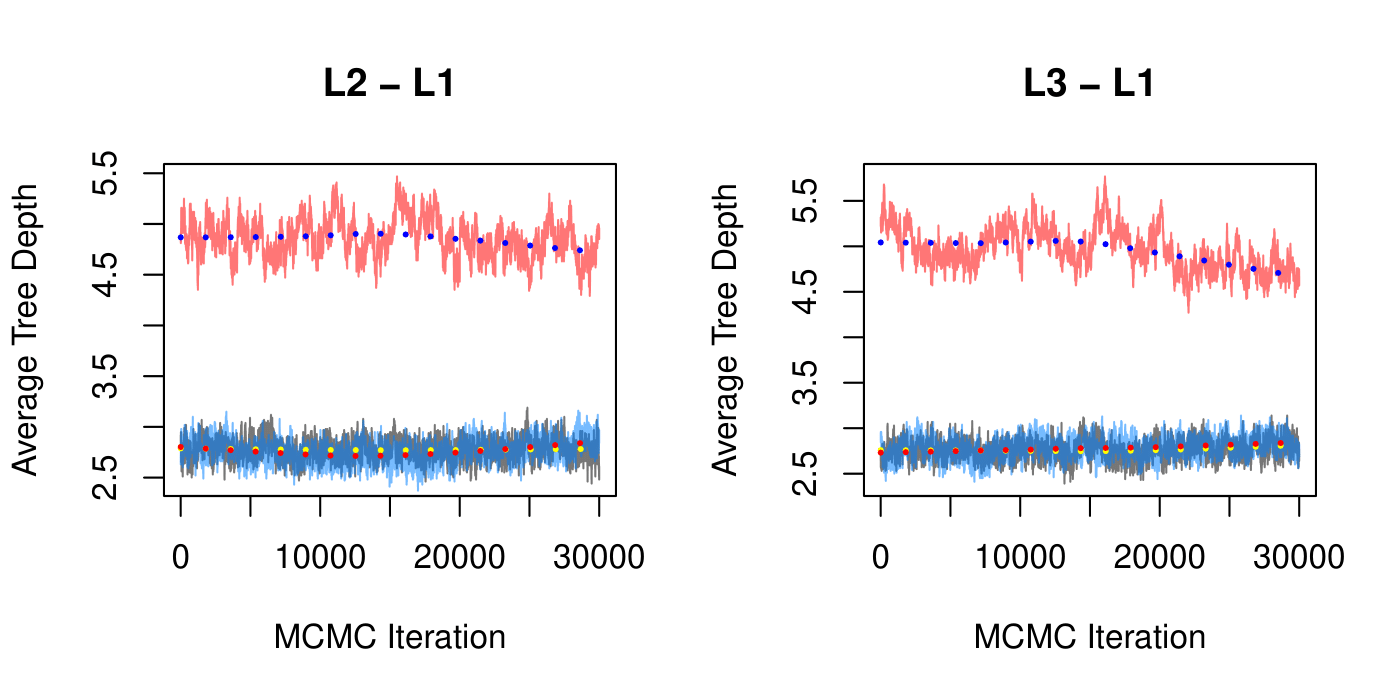}}%
	\subfloat[$\nu=C+1$, $\Psi_{12}=0$, reference level 2]{\includegraphics[scale=0.17]{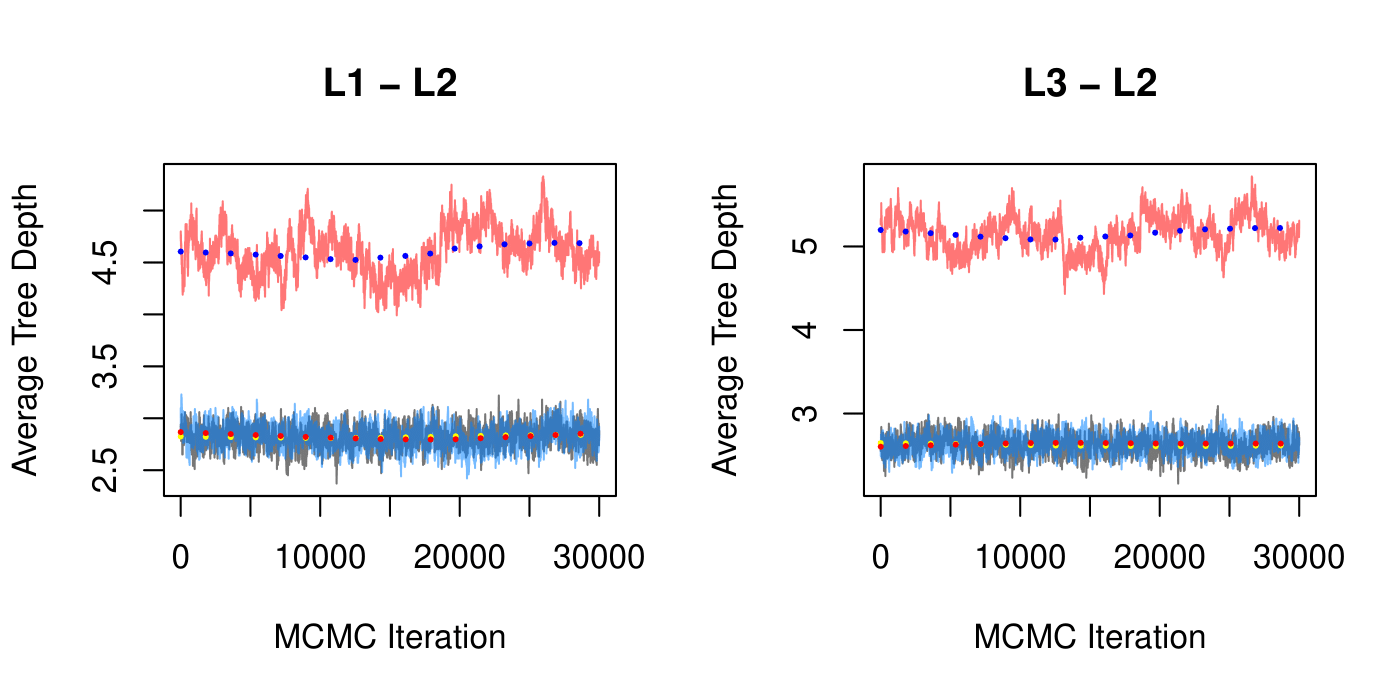}}%
	\\
	\subfloat[$\nu=C+1$, $\Psi_{12}=0$, reference level 3]{\includegraphics[scale=0.17]{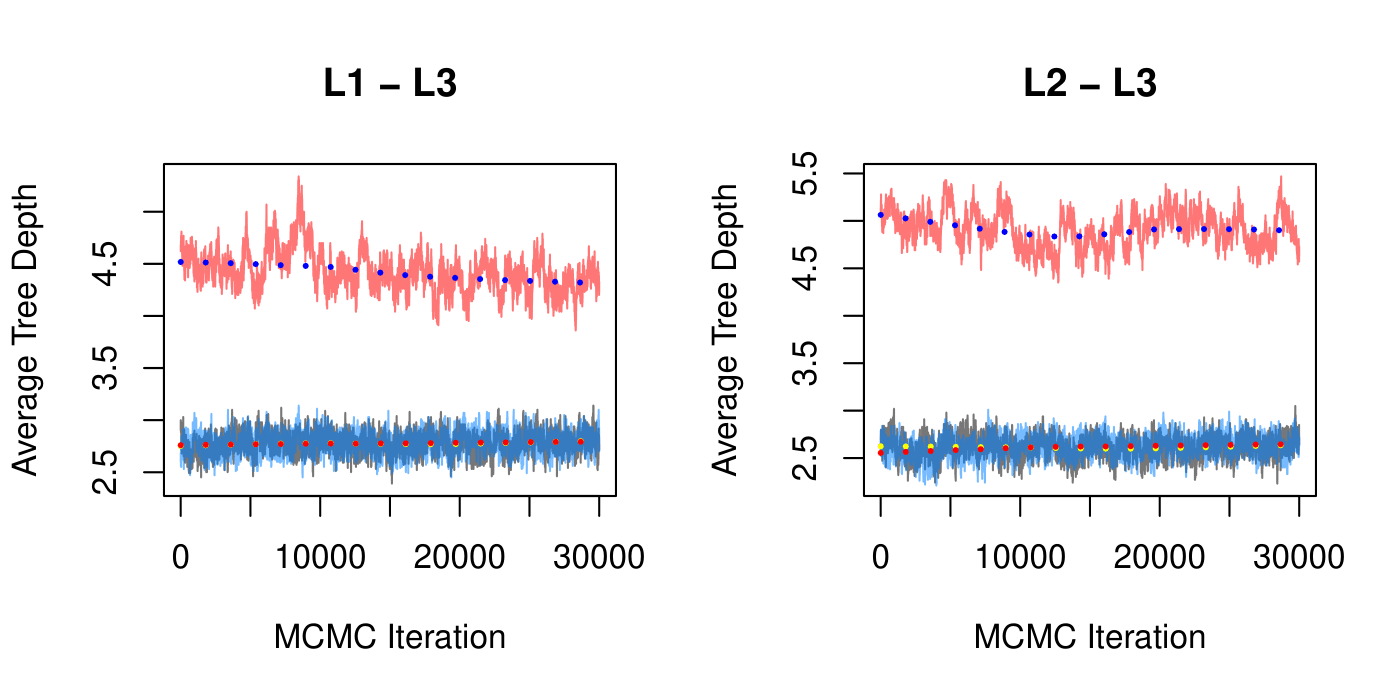}}%
	\subfloat[$\nu=C+3$, $\Psi_{12}=-0.5$, reference level 3]{\includegraphics[scale=0.17]{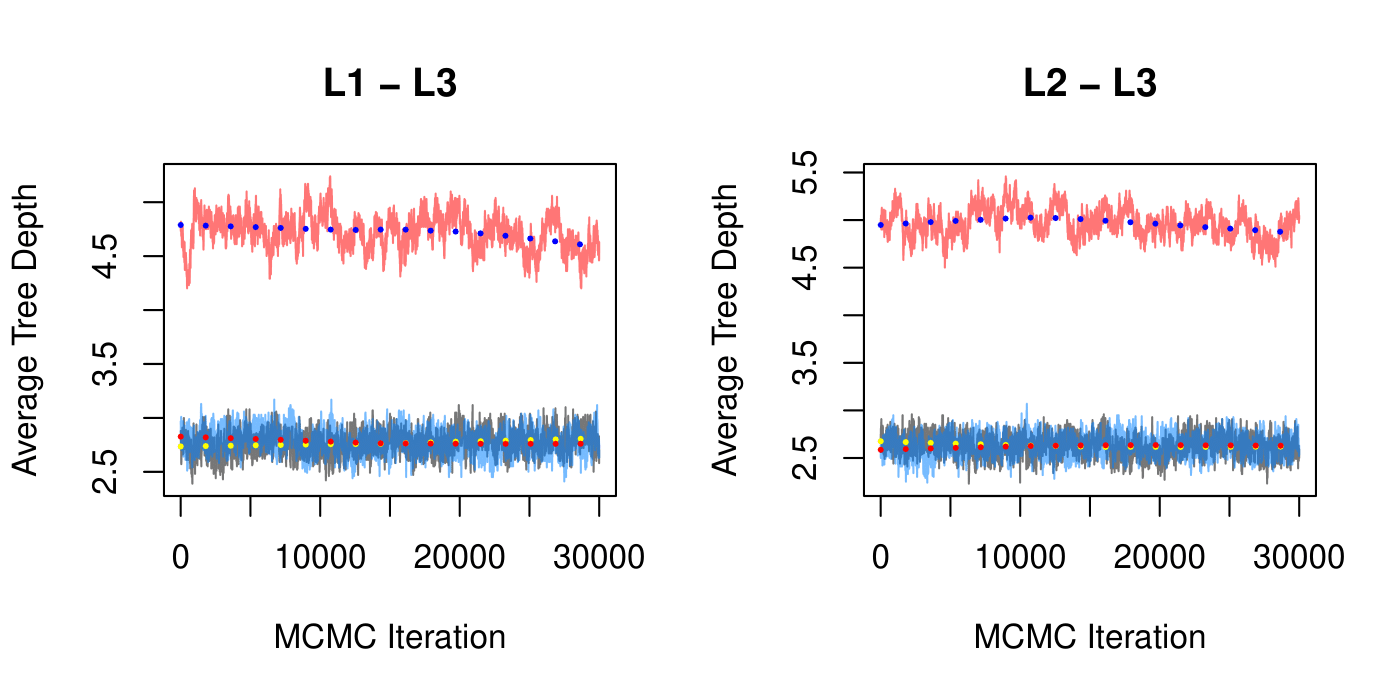}}%
	\\
	\subfloat[$\nu=C+3$, $\Psi_{12}=0.5$, reference level 3]{\includegraphics[scale=0.17]{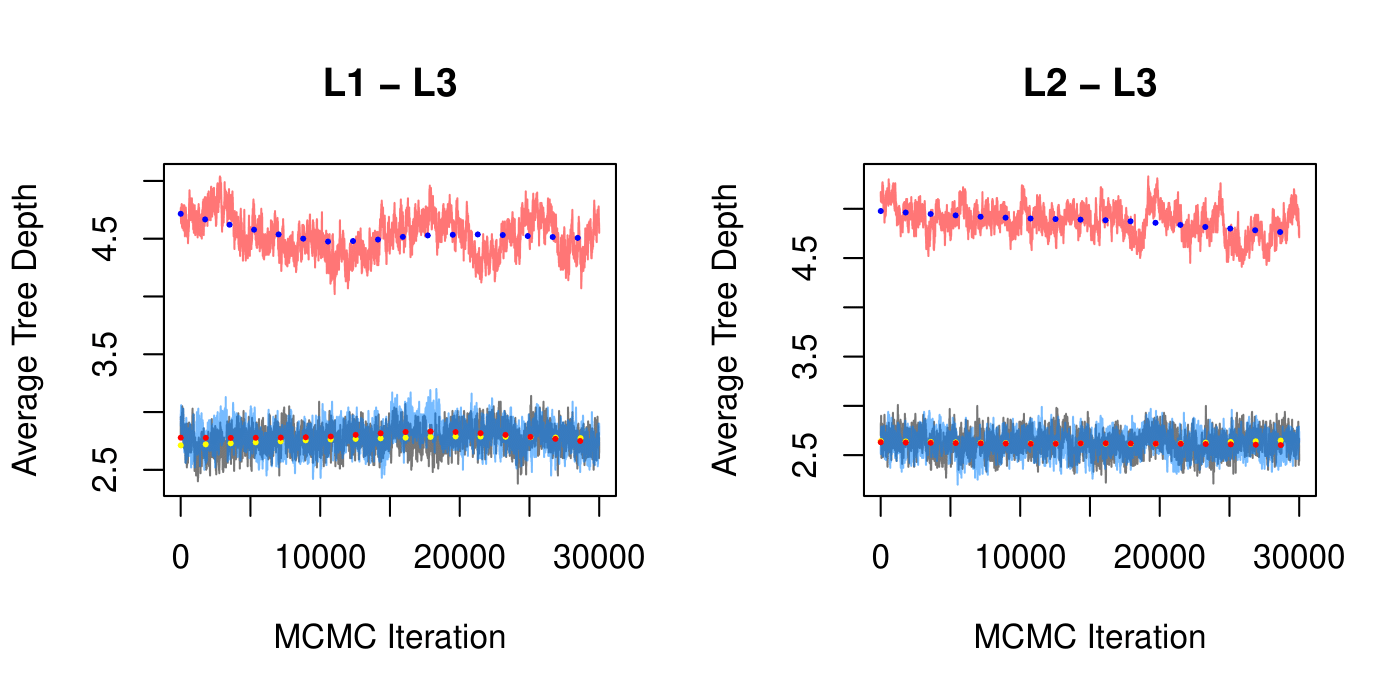}}%
	\caption{Plot of posterior average tree depth for each latent utility as time series, under simulation Setting 1 and hyperparameters described in plot labels. Red, black, and blue correspond to Algorithms [KD], [P1], and [P2], respectively. }
	\label{TDS1}
\end{figure*}

\begin{figure*}
	\centering
	\subfloat[$\nu=C+1$, $\Psi_{12}=0$]{\includegraphics[scale=0.18]{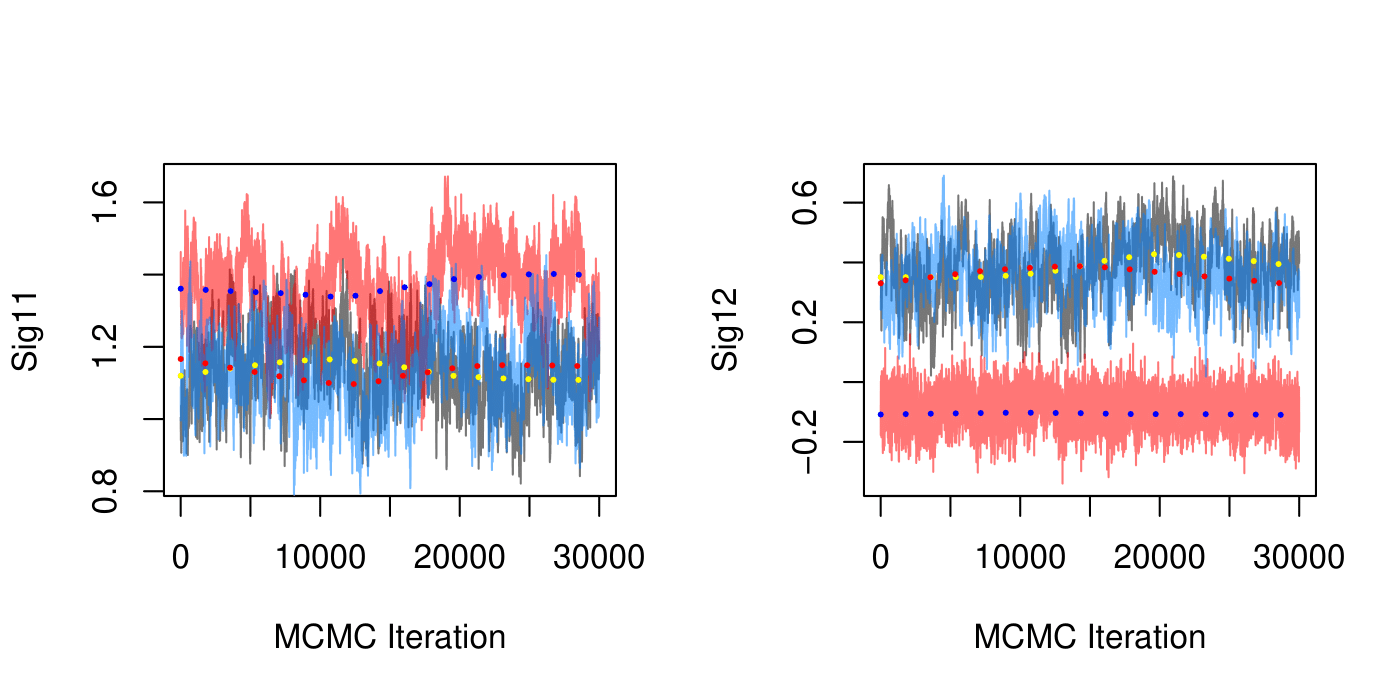}
	\includegraphics[scale=0.55]{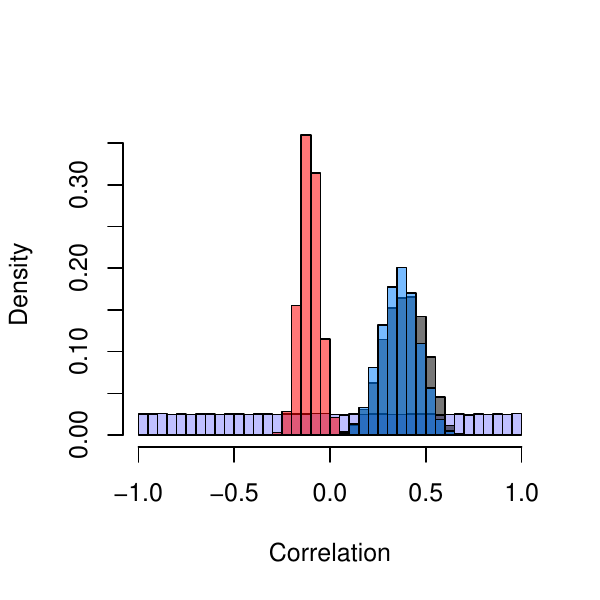}}%
	\\
	\subfloat[$\nu=C+3$, $\Psi_{12}=-0.5$]{\includegraphics[scale=0.18]{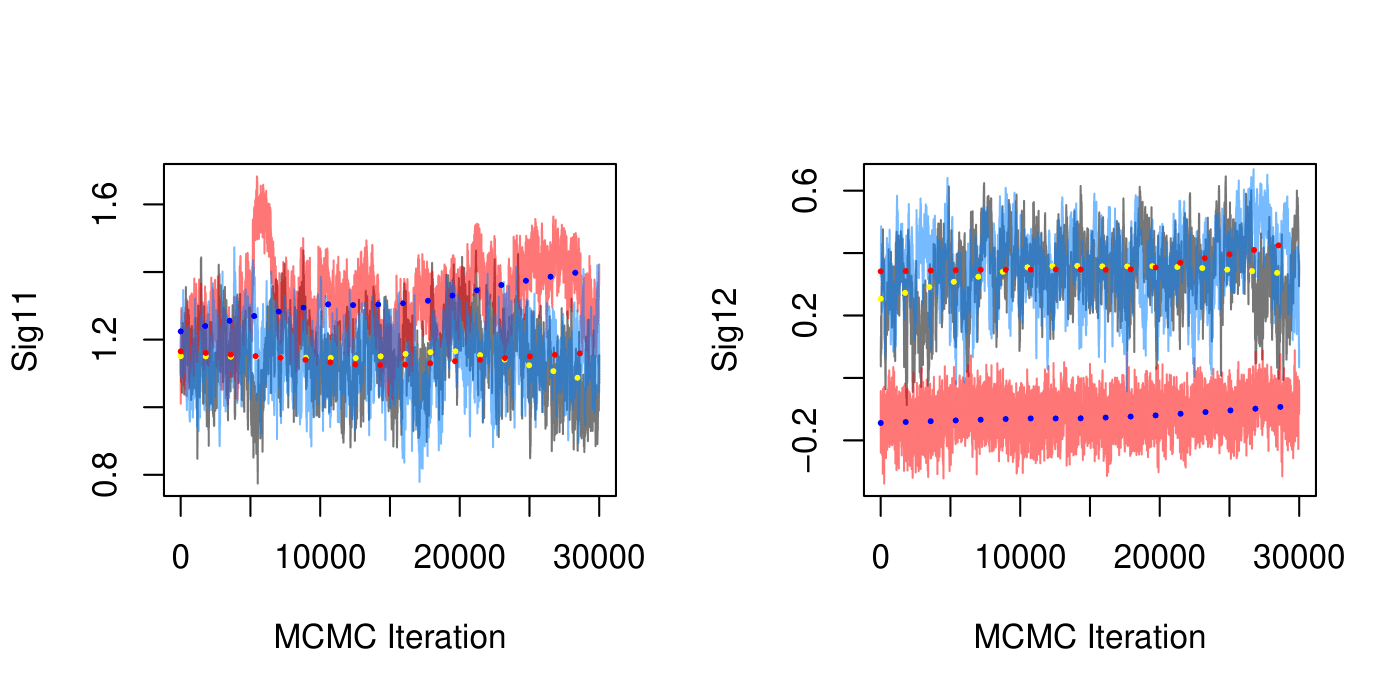}
		\includegraphics[scale=0.55]{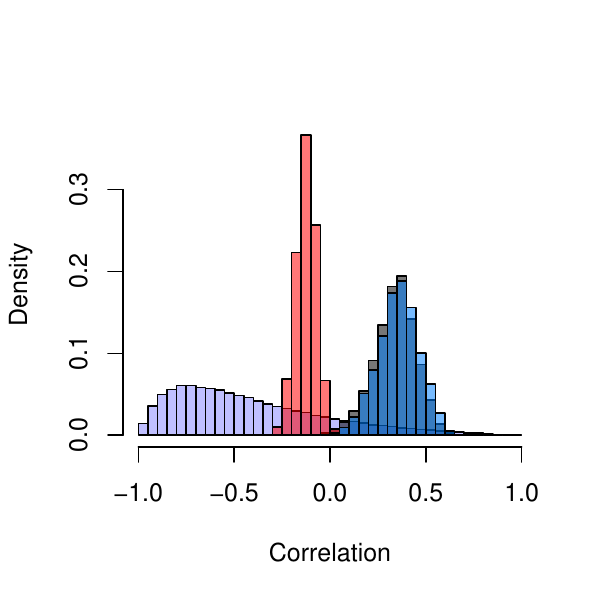}}%
	\\
	\subfloat[$\nu=C+3$, $\Psi_{12}=0.5$]{\includegraphics[scale=0.18]{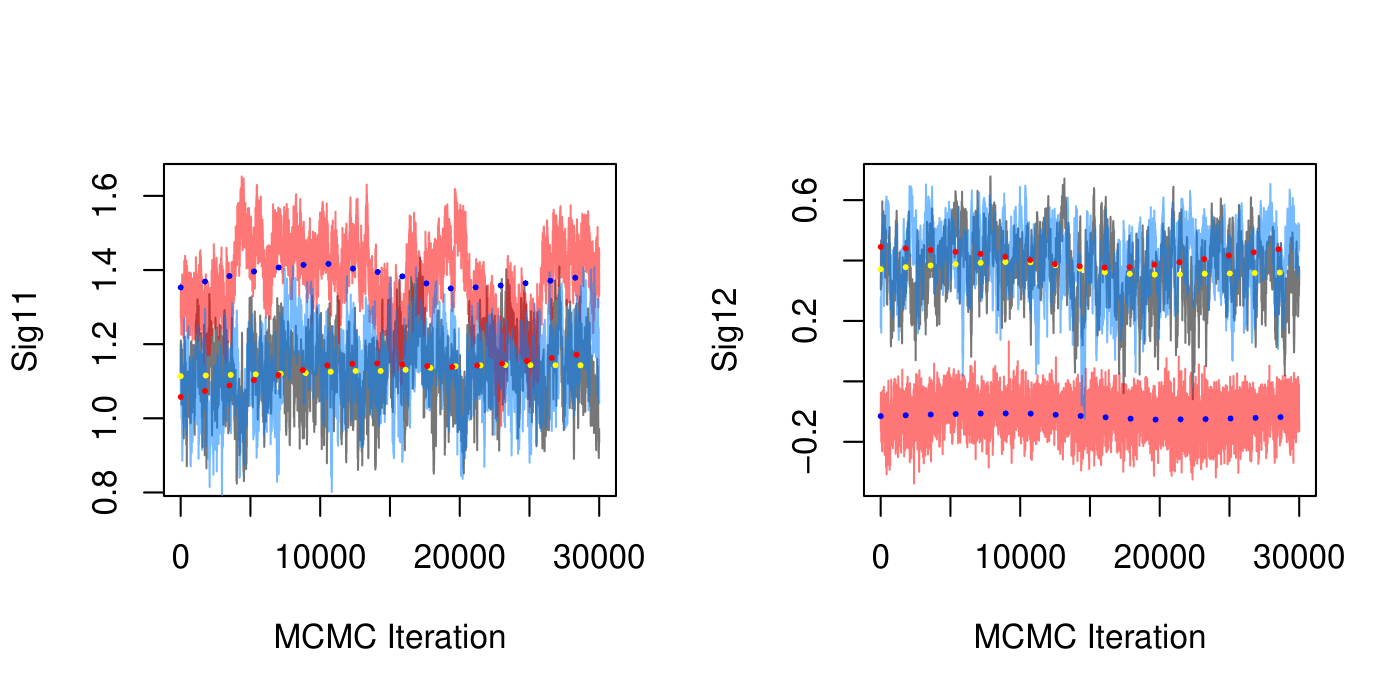}
		\includegraphics[scale=0.55]{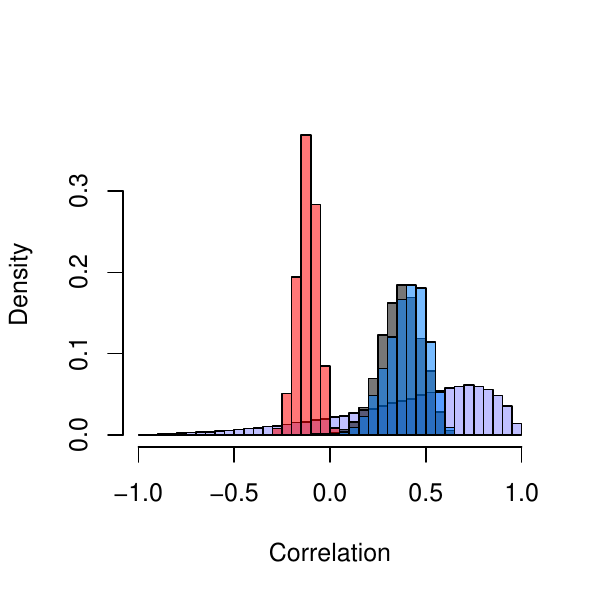}}%
	
	\caption{Plot of posterior $\sigma_{11}$ and $\sigma_{12}$ as time series on the left, and histogram of the $\sigma_{12}$ under its prior (purple), posterior from Algorithms [KD] (red), [P1] (black), and [P2] (blue); same color specification applies to the left plot. Posterior inference is conducted under Setting 1, reference level 3, and hyperparameters described in plot labels. }
	\label{SigS1}
\end{figure*}

Table \ref{Table4} shows that the three algorithms are generally robust to the specification of covariance hyperparameters. We also repeat the simulation 50 times and report the mean and standard deviation of the posterior mean of $\Sigma$ across the 50 replications (Table \ref{Table7}).
$E[\cdot|D]$ is the posterior mean based on one simulated dataset $D$. $E\{E[\cdot|D]\}$ and $S\{E[\cdot|D]\}$ are the mean and sd of $E[\cdot|D]$ across the 50 replications. The conclusions in Table \ref{Table7} mirror the conclusions in the previous results in Table \ref{Table4}. In addition, conclusions on posterior predictive accuracy from the 50 replications (Table \ref{Table8}) match those from one run of the simulation (Table \ref{Table3}). Trace plots of the MCMC convergence for the sum-of-trees (Figure \ref{TDS1}) and the covariance components (Figure \ref{SigS1}) are in Section \ref{fig}.

\subsection{Setting 2}\label{setting2}

Setting 2 is designed to investigate highly imbalanced outcomes. For the first simulated dataset under Setting 2, the frequency of outcome alternatives was $(1557,3275,168)$ and $(1601,3213,186)$ for the training and test set, respectively; the frequency of $S=3$ is less than 4$\%$, representing a very unbalanced outcome distribution. Table \ref{Table2} compares the out-of-sample accuracy of Algorithms [KD], [P1], and [P2] when the reference level is respectively set to 1, 2, and 3. The conclusions are similar to that from Table \ref{Table1}, but the sensitivity of Algorithm [KD] to the choice of reference level is even more pronounced here, while the behavior of Algorithms [P1] and [P2] remains consistent regardless of that choice.

Under reference level 3, Table \ref{Table5} shows the comparison of the out-of-sample accuracy and Table \ref{Table6} shows the posterior mean and 95 $\%$ credible intervals of the normalized covariance matrix entries $\sigma_{11}$ and $\sigma_{12}$, with an inverse Wishart prior on $\widetilde{\Sigma}$ respectively comparing a uniform ($\nu=C+1, \Psi_{12}=0$), negatively tilted ($\nu=C+3, \Psi_{12}=-0.5$), and positively tilted ($\nu=C+3, \Psi_{12}=0.5$) prior for the un-normalized correlation. Similar conclusions to Setting 1 from Tables \ref{Table3} and \ref{Table4} can be seen here, except that the deviation in the estimate of the covariance $\sigma_{12}$ between Algorithm [KD] and Algorithms [P1] and [P2] is even more extreme in this case with the outcome unbalanced. As mentioned in Section \ref{setting1}, the conditional covariance between $W_1$ and $W_2$ is 0.5, and the estimated posterior mean of covariance from Algorithms [P1] and [P2] taking values near 0.8, which agrees with the true correlation in sign. Algorithm [KD] yields negative estimates around -0.35 and its impact on predicted value accuracy is similarly poor as in Setting 1. 

\begin{figure*}
	\centering
	\subfloat[$\nu=C+1$, $\Psi_{12}=0$, reference level 1]{\includegraphics[scale=0.17]{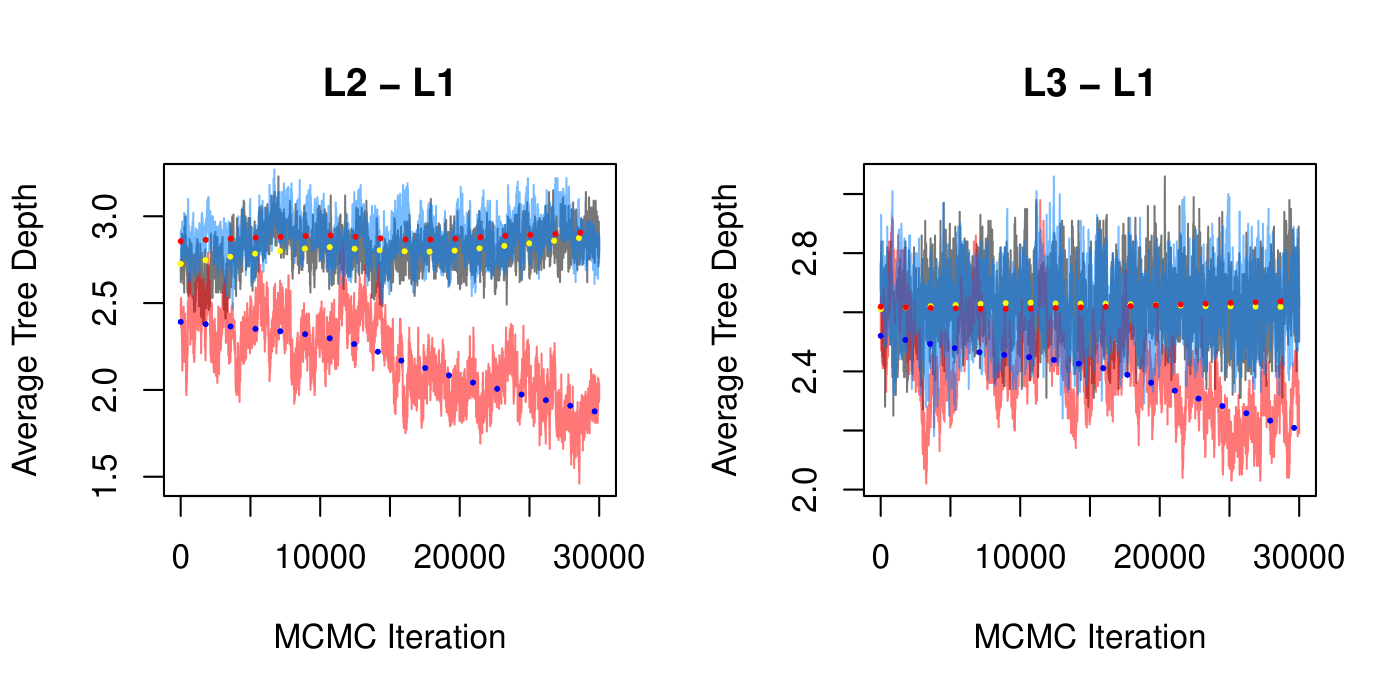}}%
	\subfloat[$\nu=C+1$, $\Psi_{12}=0$, reference level 2]{\includegraphics[scale=0.17]{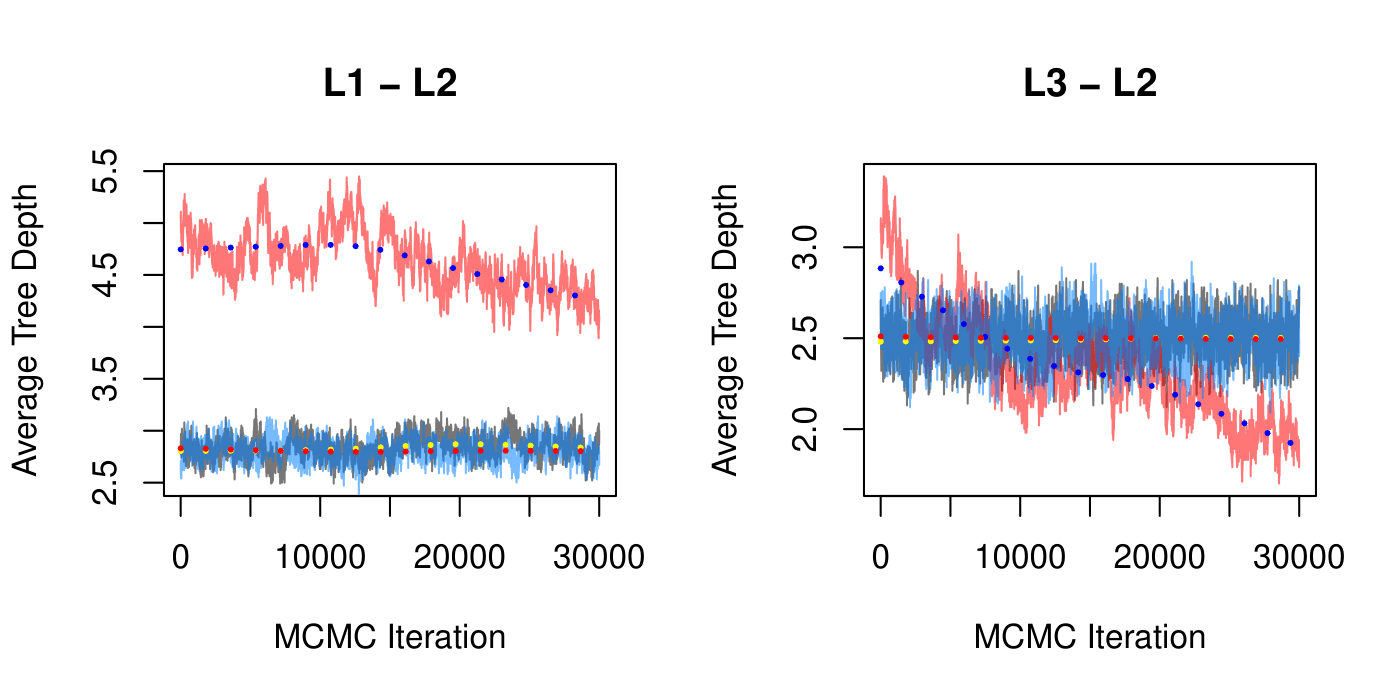}}%
	\\
	\subfloat[$\nu=C+1$, $\Psi_{12}=0$, reference level 3]{\includegraphics[scale=0.17]{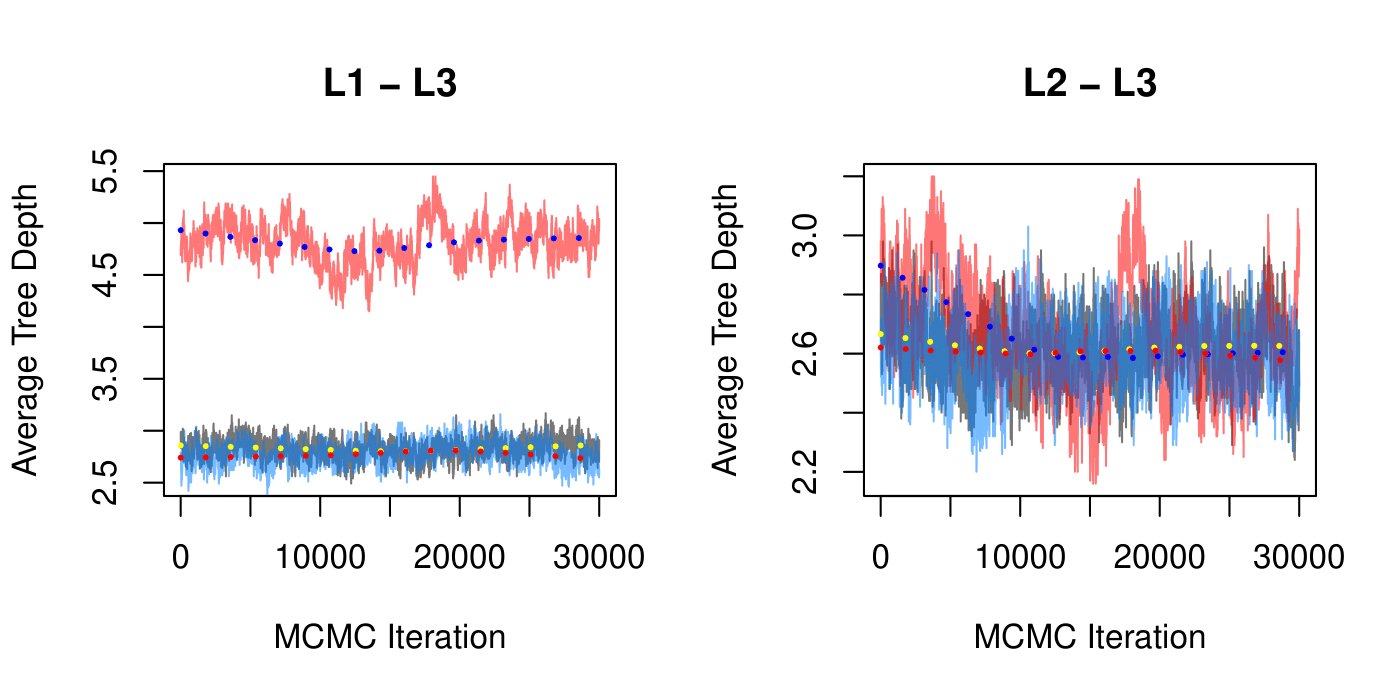}}%
	\subfloat[$\nu=C+3$, $\Psi_{12}=-0.5$, reference level 3]{\includegraphics[scale=0.17]{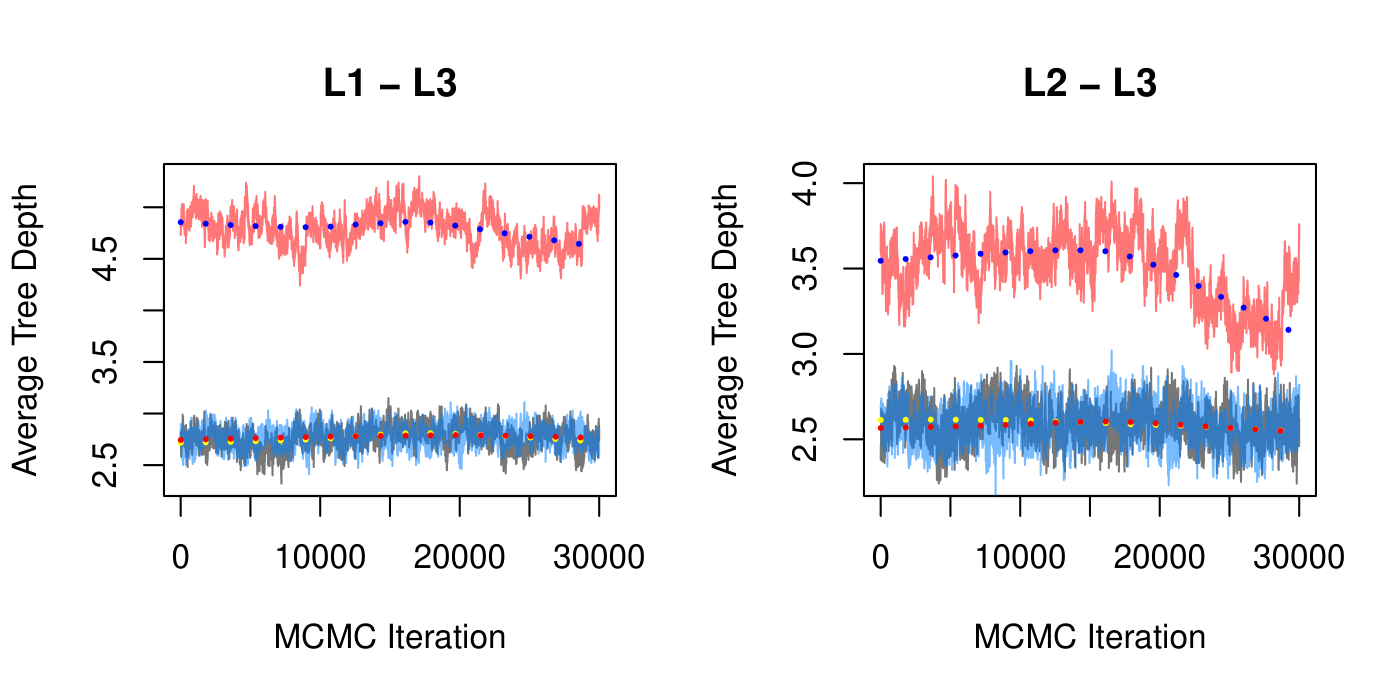}}%
	\\
	\subfloat[$\nu=C+3$, $\Psi_{12}=0.5$, reference level 3]{\includegraphics[scale=0.17]{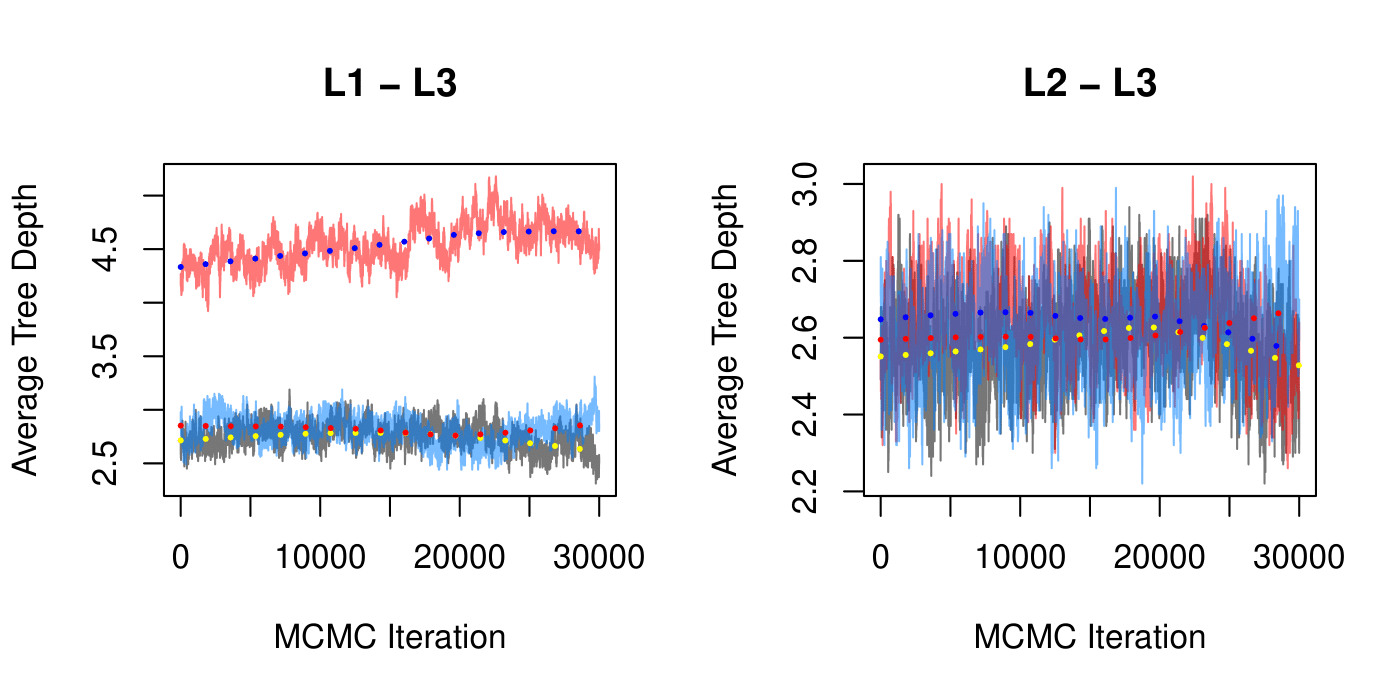}}%
	\caption{Plot of posterior average tree depth for each latent utility as time series, under simulation Setting 2 and hyperparameters described in plot labels. Red, black, and blue correspond to Algorithms [KD], [P1], and [P2], respectively. }
	\label{TDS2}
\end{figure*}

\begin{figure*}
	\centering
	\subfloat[$\nu=C+1$, $\Psi_{12}=0$]{\includegraphics[scale=0.18]{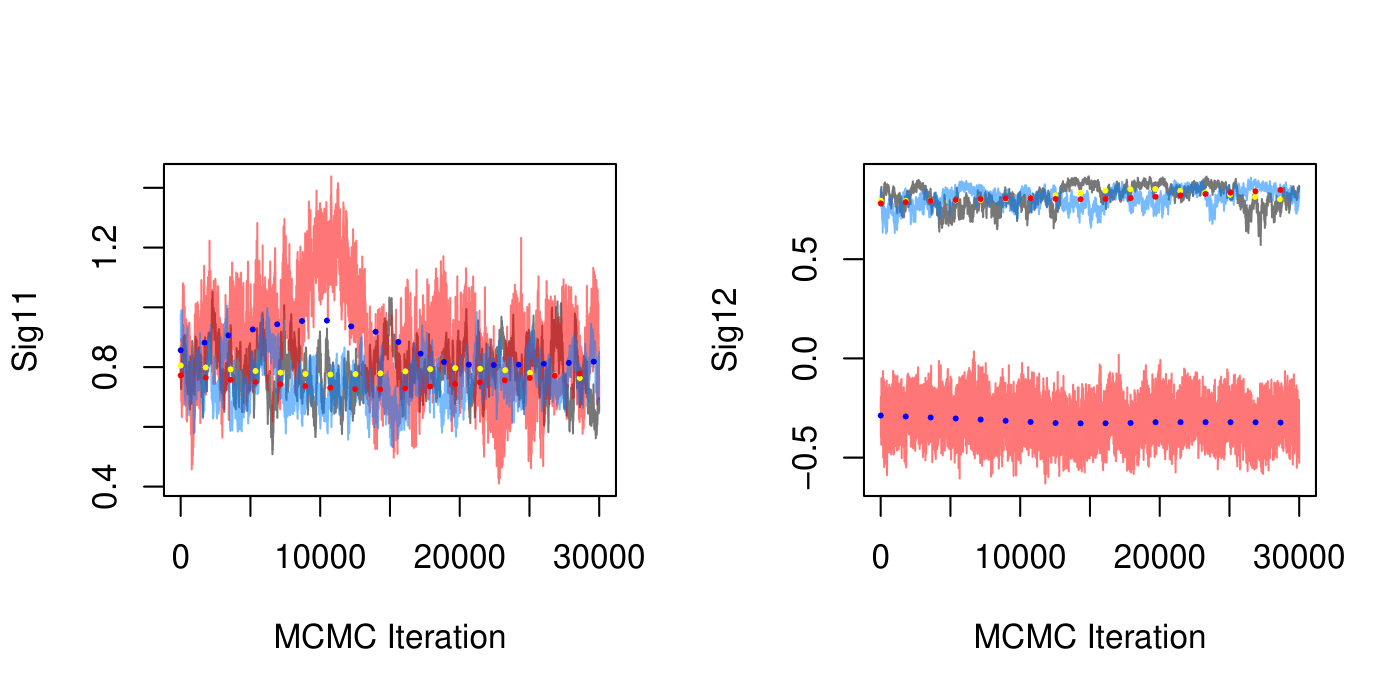}
		\includegraphics[scale=0.55]{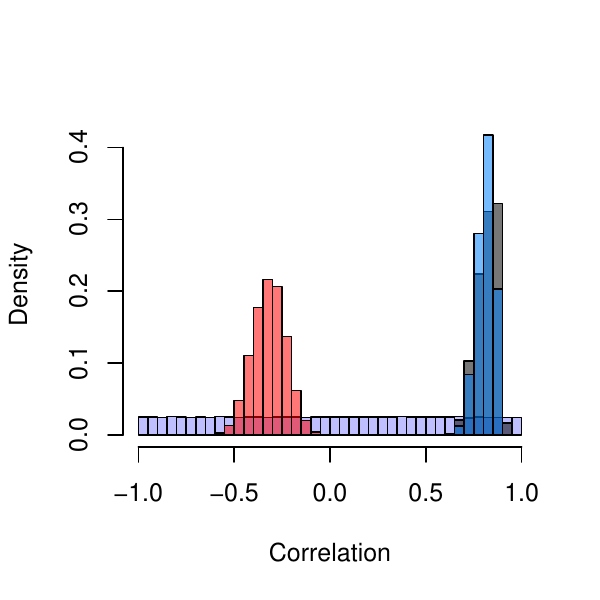}}%
	\\
	\subfloat[$\nu=C+3$, $\Psi_{12}=-0.5$]{\includegraphics[scale=0.18]{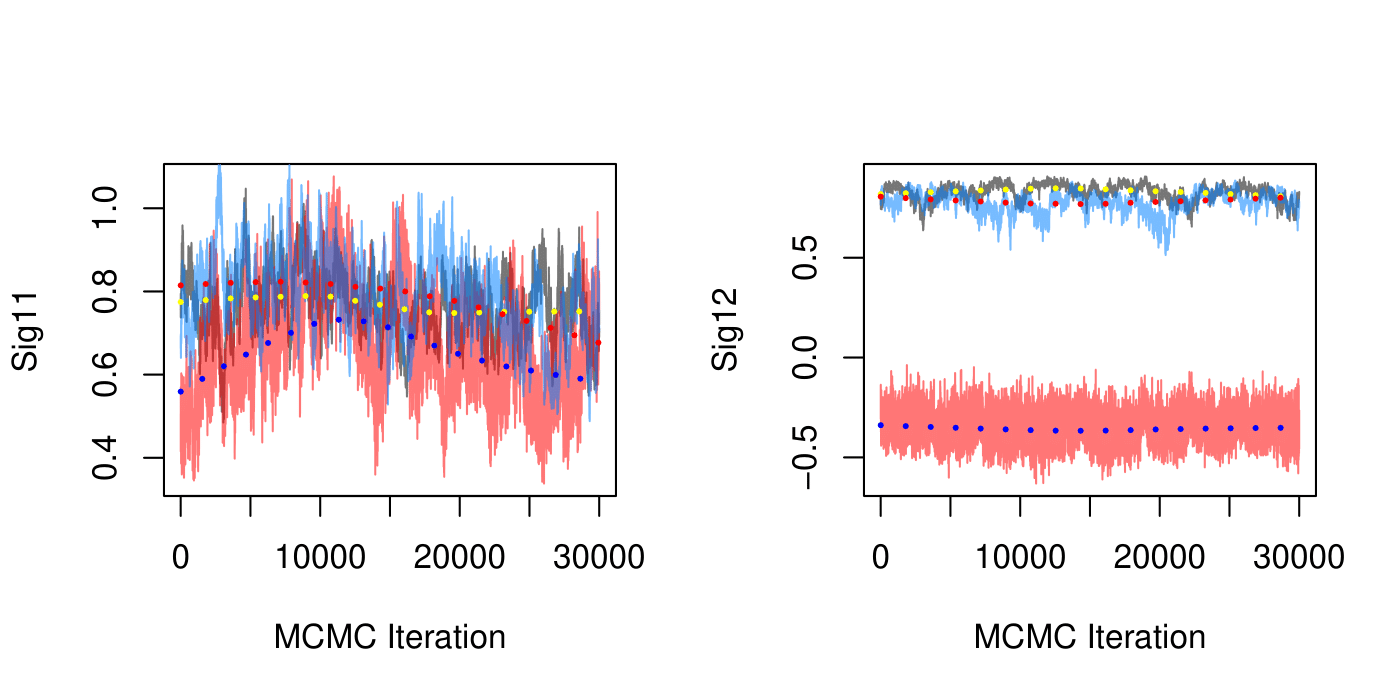}
		\includegraphics[scale=0.55]{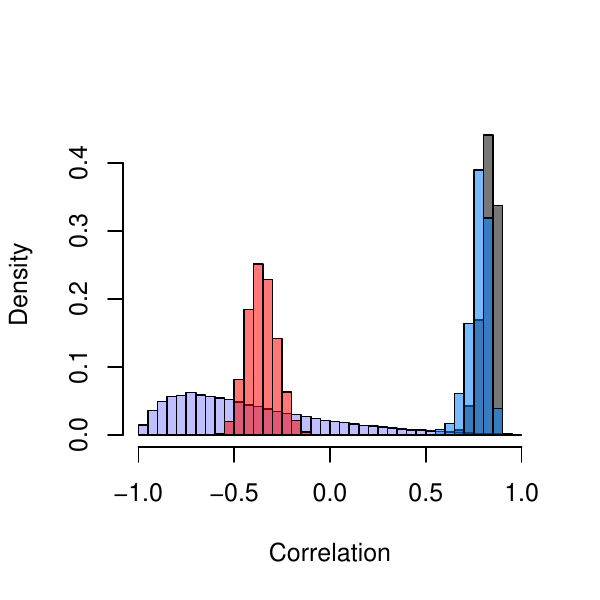}}%
	\\
	\subfloat[$\nu=C+3$, $\Psi_{12}=0.5$]{\includegraphics[scale=0.18]{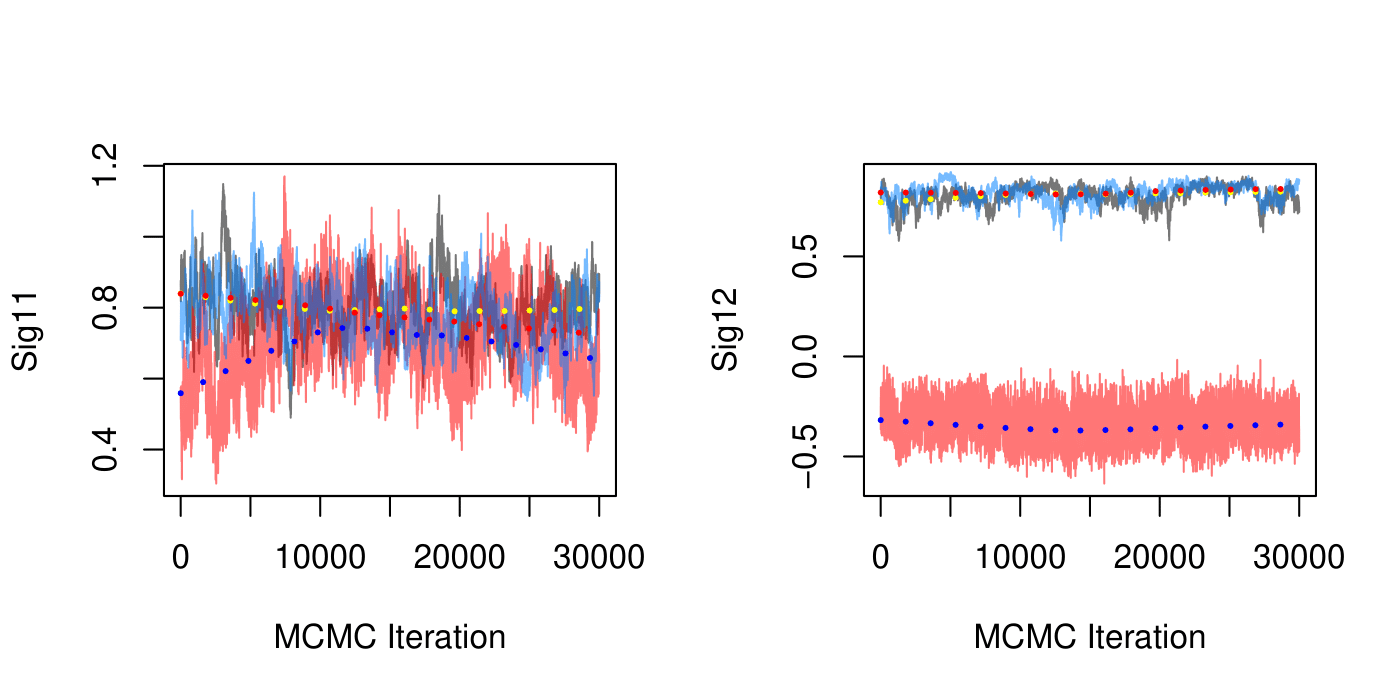}
		\includegraphics[scale=0.55]{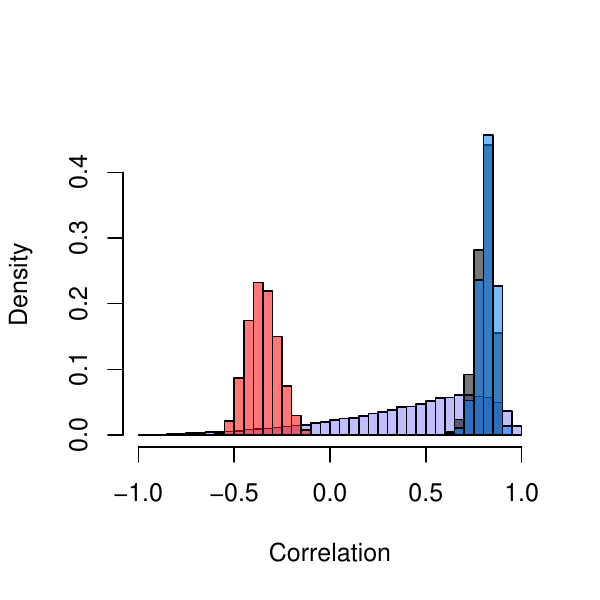}}%
	
	\caption{Plot of posterior $\sigma_{11}$ and $\sigma_{12}$ as time series on the left, and histogram of the $\sigma_{12}$ under its prior (blue), posterior from Algorithms [KD] (red), [P1] (black), and [P2] (blue). Posterior inference is conducted under simulation Setting 2, reference level 3, and hyperparameters described in plot labels. Red, black, and blue correspond to Algorithms [KD], [P1], and [P2], respectively. }
	\label{SigS2}
\end{figure*}

Table \ref{Table9} and \ref{Table10} report the posterior mean of $\Sigma$ and the posterior accuracy, by their mean and standard deviation across 50 replications. As before, the observations are consistent with those under one run of simulation. Moreover, Figure \ref{TDS2} and \ref{SigS2} provide diagnostic plots of the MCMC convergence for the sum-of-trees and the covariance components, respectively; it is obvious that Algorithms [P1] and [P2] converge faster than Algorithm [KD]. When the outcome is highly imbalanced, posterior convergence is more difficult than the balanced case. Compared to Algorithms [P1] and [P2], the approach Algorithm [KD] uses to update the sum-of-trees component using un-normalized latent utilities may further obstruct posterior convergence.   

The simulation study in Sections \ref{setting1} and \ref{setting2} shows that: (a) the proposals, Algorithms [P1] and [P2], have better posterior predictive accuracy than Algorithm [KD] in general; (b) the proposals are more robust than Algorithm [KD] such that the degree of accuracy is unaffected by choices of reference level, imbalance in the categorical distribution, and the choice of prior specifications in the sum-of-trees and variance components; (c) sign of the posterior estimates for the covariance parameter agrees with the underlying truth under the proposals but not under Algorithm [KD].

\section{Application}\label{app2}

In this application, we investigate patients' retention in HIV care after enrollment as a function of their baseline characteristics and treatment status. The data were extracted from electronic health records of adults enrolled in HIV care between June 1st 2008 and August 23rd 2016 in AMPATH, an HIV care monitoring program in Kenya. We look at a 200-days window after the initial care encounter and split the data into training and test sets of sample sizes 49,942 and 26,714, respectively. We define the outcome as disengagement, engagement, and reported death, where engagement in care means there was at least one visit to the clinic for HIV care during the first 200 days after a patient's initial encounter, and disengagement otherwise if the person was not reported dead. The outcome distribution is extremely imbalanced, such that the frequency of disengagement, engagement, and death is 16$\%$, 80$\%$, and 4$\%$, respectively. Covariates include baseline age, gender, year of enrollment, travel time to clinic, marriage status, weight, height, baseline treatment status, indication of CD4 measurement at or post baseline, and the most recent CD4. Table \ref{Table11} summarizes the observed distribution of  each covariate stratified by outcome level. 

We use 10,000 posterior draws after a burn-in of 10,000 and keep other settings the same as in simulations. Table \ref{Table12} compares the posterior accuracy for Algorithms [KD], [P1], and [P2]. Algorithm [KD] has posterior mode accuracy close to, but not as good as, that from Algorithms [P1] and [P2]. In terms of posterior agreement accuracy, Algorithm [KD] is substantially inferior to the proposals, indicating that the proposals yield a better separation between the random utility for the true outcome level and those for the other outcome alternatives. In terms of the stability in accuracy measures with respect to the choice of reference level, the proposals also performed better than Algorithm [KD].

\begin{figure*}
	\centering
	\subfloat[Reference level 1 (disengagement)]{\includegraphics[scale=0.6]{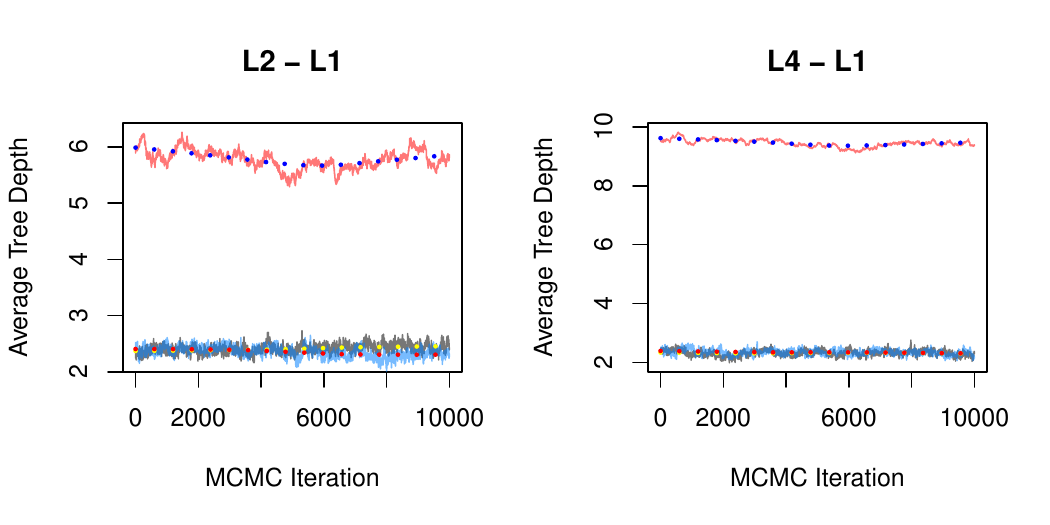}
	\includegraphics[scale=0.55]{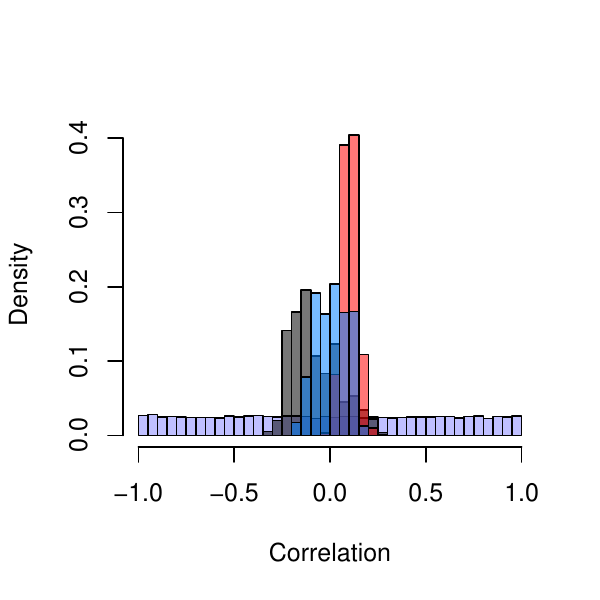}}%
	\\
	\subfloat[Reference level 2 (engagement)]{\includegraphics[scale=0.6]{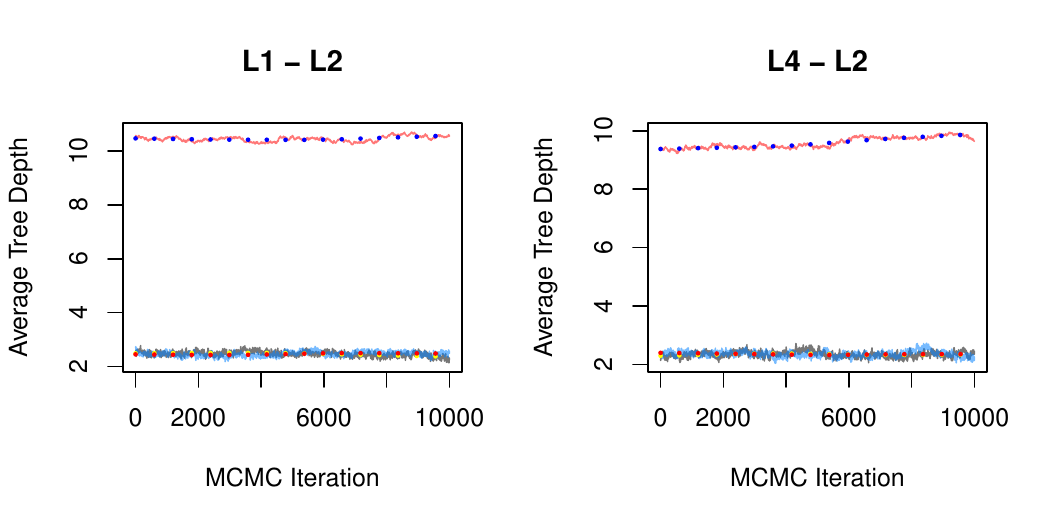}
	\includegraphics[scale=0.55]{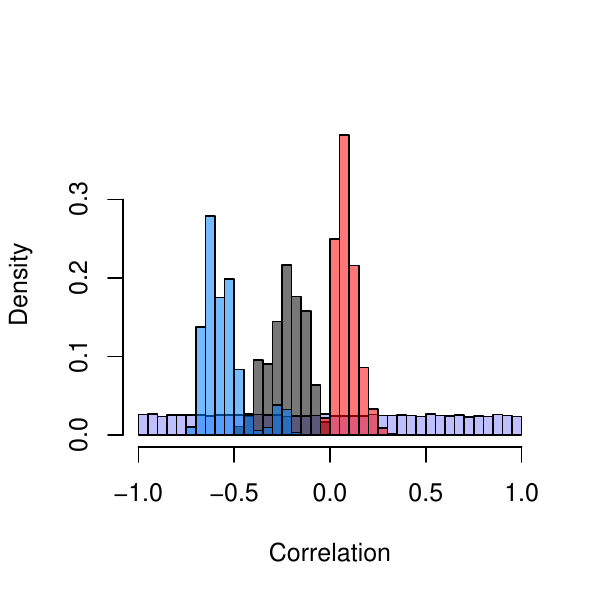}}%
	\\
	\subfloat[Reference level 4 (death)]{\includegraphics[scale=0.6]{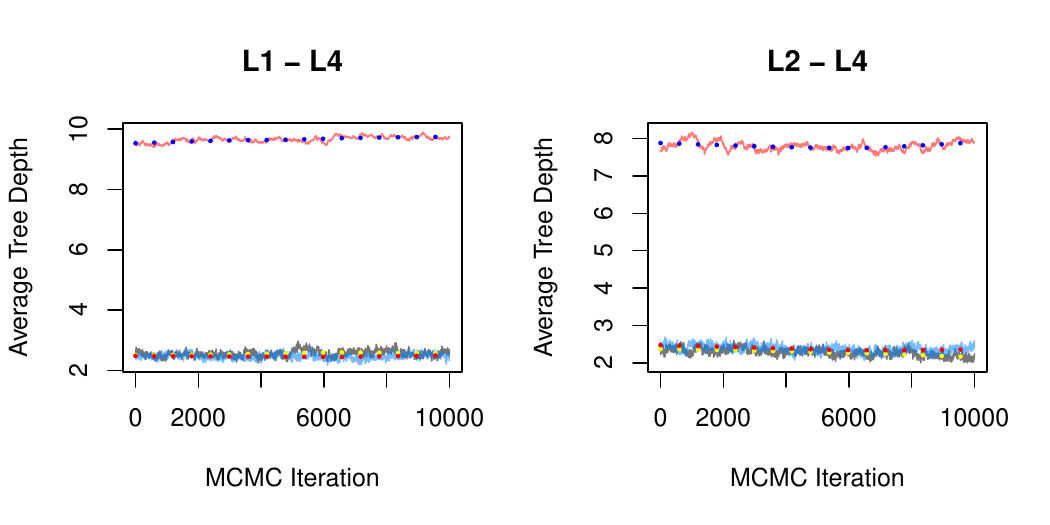}
	\includegraphics[scale=0.55]{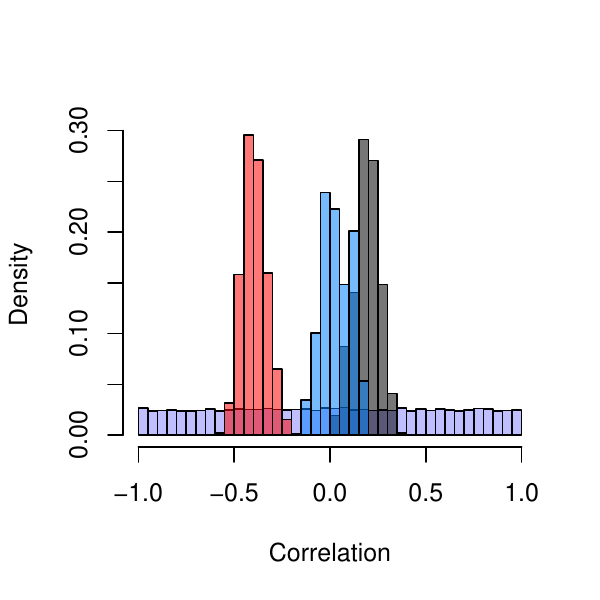}}%
	
\caption{Plot of posterior average tree depth for each latent utility as time series for the application analysis on AMPATH data on the left, and histogram of the $\sigma_{12}$ under its prior (purple), posterior from Algorithms [KD] (red), [P1] (black), and [P2] (blue); same color specification applies to the left plot. Posterior inference is under $\nu=C+1$, $\Psi_{12}=0$, with reference level as indicated in plot labels.}
	\label{TDApp}
\end{figure*}

Under the reference level being disengagement, the first row of Figure \ref{TDApp} presents the MCMC convergence plots of the average tree depth corresponding to latent variables $W_1=Z_{eng}-Z_{diseng}$ and  $W_2=Z_{death}-Z_{diseng}$, and the histogram of the posteriors of $\sigma_{12} = \text{Cov}(W_1, W_2)$, where $(Z_{eng}, Z_{diseng}, Z_{death})$ are latent utilities corresponding to each of the outcome levels and $\sigma_{12}$ is the normalized conditional covariance of $W_1$ and $W_2$. The plots show that the average tree depths are around 6 and 9 respectively for $W_1$ and $W_2$ under Algorithm [KD], and approximately 2 for those under Algorithms [P1] and [P2]. The Bayesian regularization priors that favor shallow trees do not work well for Algorithm [KD], as a tree depth of 6 allows up to $2^6$ leaves, which increases the risk of over-fitting and makes the stochastic tree search inefficient. The second and third rows of Figure \ref{TDApp} set engagement and reported death as the reference level, respectively, and the latent variables are defined accordingly. Similar conclusions are observed for tree depth. Under all choices of the reference level, the histogram of $\sigma_{12}$ from Algorithms [P1] and [P2] agree on the sign of $\sigma_{12}$, which was demonstrated in previous simulations to match the sign of the true value of the underlying $\sigma_{12}$.

\section{Concluding Remarks}\label{last2}

While computational performance is an important criterion in building Gibbs sampler for complicated models, the dependency structure and sampling schemes are as crucial for devising an algorithm that generates a Markov chain with computational efficiency and fast mixing rates. We explore the data augmentation scheme for MPBART involved in KD, and propose two alternative algorithms for MPBART that have improved computational and theoretical properties. Theoretically, we prove that the mixing rate of our proposals (Algorithms [P1] and [P2]) is at least as good as Algorithm [KD] for both the mean and covariance matrix of the latent variables.

To assess computational performance, we consider two aspects, predictive accuracy and posterior convergence. In terms of predictive accuracy, we point out the limitation of the classification error metric in \citep{kindo_multinomial_2016}, and also investigate the posterior agreement accuracy. Through numerical studies, we observe that Algorithm [KD] performs well under the posterior mode metric but its variation across posterior predictions tends to be too large. Our proposals yield better predictive accuracy than Algorithm [KD], with the improvement substantial under the posterior agreement accuracy.  

We also observed that for Algorithm [KD], posterior trees on average converge to more complex trees than that in our proposals, and this has two important implications. First, the Bayesian regularization priors from BART which put more weight on smaller trees do not seem to work well under Algorithm [KD]. Second, the convergence of the sum-of-trees parameters has a large influence not only on the posterior predictive accuracy but also on the convergence of the covariance matrix. As shown in simulations, the posterior covariance in Algorithm [KD] may seem to converge but often to a value that has the wrong sign. In Appendix \ref{Sbysig12} we further explore how the correlation of the latent utilities affects the outcome distribution, and show that an estimated covariance of the wrong sign can lead to a sum-of-trees component with values systematically different from the true data generating mechanism.

%{\noindent \em Remainder omitted in this sample. See http://www.jmlr.org/papers/ for full paper.}

% Acknowledgements should go at the end, before appendices and references


\newpage

\section*{Tables}

\begin{table}[H]%Table 1 in BartGcomp_19-11-26_Tables.R
	\centering
\begin{tabular}{ccccccc}
	& \multicolumn{6}{c}{Posterior Agreement Accuracy}                                                                                             \\ \hline
	& \multicolumn{3}{c}{Train}                                          & \multicolumn{3}{c}{Test}                                           \\ \hline
	Ref Level            & KD                 & P1                 & P2                 & KD                 & P1                 & P2                 \\ \hline
	1                    & 0.57                 & 0.89                 & 0.89                 & 0.54                 & 0.87                 & 0.87                 \\
	2                    & 0.60                 & 0.89                 & 0.90                 & 0.58                 & 0.88                 & 0.88                 \\
	3                    & 0.60                 & 0.90                 & 0.90                 & 0.58                 & 0.88                 & 0.88                 \\ \hline
	\multicolumn{1}{l}{} & \multicolumn{1}{l}{} & \multicolumn{1}{l}{} & \multicolumn{1}{l}{} & \multicolumn{1}{l}{} & \multicolumn{1}{l}{} & \multicolumn{1}{l}{} \\
	& \multicolumn{6}{c}{Posterior Mode Accuracy}                                                                                             \\ \hline
	& \multicolumn{3}{c}{Train}                                          & \multicolumn{3}{c}{Test}                                           \\ \hline
	Ref Level            & KD                 & P1                 & P2                 & KD                 & P1                 & P2                 \\ \hline
	1                    & 0.93                 & 0.95                 & 0.94                 & 0.87                 & 0.91                 & 0.92                 \\
	2                    & 0.92                 & 0.94                 & 0.94                 & 0.87                 & 0.92                 & 0.91                 \\
	3                    & 0.92                 & 0.94                 & 0.95                 & 0.87                 & 0.92                 & 0.92                 \\ \hline
\end{tabular}
\caption{Accuracy comparison of Algorithms [KD], [P1], and [P2]. Training and test datasets each with 5000 observations are generated under Setting 1 with reference level 3. Posterior predictive accuracy measured by \eqref{accmean} and \eqref{accmode} are reported under reference levels 1, 2, and 3. The prior of $\widetilde{\Sigma}$ is $\text{Inv-Wishart}(C+1,I_C)$, where $C=2$.}
\label{Table1}
\end{table}

\begin{table}[H]%Table 2 in BartGcomp_19-11-26_Tables.R
	\centering
\begin{tabular}{ccccccc}
	& \multicolumn{6}{c}{Posterior Agreement Accuracy}                                                                                             \\ \hline
	& \multicolumn{3}{c}{Train}                                          & \multicolumn{3}{c}{Test}                                           \\ \hline
	Ref Level            & KD                 & P1                 & P2                 & KD                 & P1                 & P2                 \\ \hline
	1                    & 0.60                 & 0.90                 & 0.90                 & 0.59                 & 0.87                 & 0.87                 \\
	2                    & 0.57                 & 0.90                 & 0.90                 & 0.55                 & 0.88                 & 0.88                 \\
	3                    & 0.65                 & 0.91                 & 0.90                 & 0.63                 & 0.88                 & 0.88                 \\ \hline
	\multicolumn{1}{l}{} & \multicolumn{1}{l}{} & \multicolumn{1}{l}{} & \multicolumn{1}{l}{} & \multicolumn{1}{l}{} & \multicolumn{1}{l}{} & \multicolumn{1}{l}{} \\
	& \multicolumn{6}{c}{Posterior Mode Accuracy}                                                                                             \\ \hline
	& \multicolumn{3}{c}{Train}                                          & \multicolumn{3}{c}{Test}                                           \\ \hline
	Ref Level            & KD                 & P1                 & P2                 & KD                 & P1                 & P2                 \\ \hline
	1                    & 0.90                 & 0.95                 & 0.94                 & 0.87                 & 0.91                 & 0.91                 \\
	2                    & 0.93                 & 0.95                 & 0.95                 & 0.88                 & 0.91                 & 0.91                 \\
	3                    & 0.92                 & 0.96                 & 0.95                 & 0.89                 & 0.91                 & 0.91                 \\ \hline
\end{tabular}
	\caption{Accuracy comparison of Algorithms [KD], [P1], and [P2]. Training and test datasets each with 5000 observations are generated under Setting 2 with reference level 3. Posterior predictive accuracy measured by \eqref{accmean} and \eqref{accmode} are reported under reference levels 1, 2, and 3. The prior of $\widetilde{\Sigma}$ is $\text{Inv-Wishart}(C+1,I_C)$, where $C=2$.}
	\label{Table2}
\end{table}

\begin{table}[H]%Table 3 in BartGcomp_19-11-26_Tables.R
	\centering

\begin{tabular}{cccccccc}
	&      & \multicolumn{6}{c}{Posterior Agreement Accuracy}          \\ \hline
	&      & \multicolumn{3}{c}{Train} & \multicolumn{3}{c}{Test} \\ \hline
	$\nu - C$ & $\Psi_{12}$    & KD    & P1   & P2   & KD    & P1   & P2  \\ \hline
	1  & 0    & 0.60    & 0.90   & 0.90   & 0.59    & 0.87   & 0.87  \\
	3  & -0.5 & 0.57    & 0.90   & 0.90   & 0.55    & 0.88   & 0.88  \\
	3  & 0.5  & 0.65    & 0.91   & 0.90   & 0.63    & 0.88   & 0.88  \\ \hline
	&      &         &        &        &         &        &       \\
	&      & \multicolumn{6}{c}{Posterior Mode Accuracy}          \\ \hline
	&      & \multicolumn{3}{c}{Train} & \multicolumn{3}{c}{Test} \\ \hline
	$\nu - C$ & $\Psi_{12}$    & KD    & P1   & P2   & KD    & P1   & P2  \\ \hline
	1  & 0    & 0.90    & 0.95   & 0.94   & 0.87    & 0.91   & 0.91  \\
	3  & -0.5 & 0.93    & 0.95   & 0.95   & 0.88    & 0.91   & 0.91  \\
	3  & 0.5  & 0.92    & 0.96   & 0.95   & 0.89    & 0.91   & 0.91  \\ \hline
\end{tabular}
	\caption{Accuracy comparison of Algorithms [KD], [P1], and [P2] under reference level 3. Training and test datasets each with 5000 observations are generated under Setting 1 with reference level 3. The prior of $\widetilde{\Sigma}$ is $\text{Inv-Wishart}(\nu,\Psi)$, where $\Psi_{11} = \Psi_{22} = 1$. Posterior predictive accuracy measured by \eqref{accmean} and \eqref{accmode} are reported under $(\nu-C,\Psi_{12})$ being $(1,0)$, $(3,-0.5)$, and $(3,0.5)$. }
	\label{Table3}
\end{table}

\begin{table}[H]%Table 4 in BartGcomp_19-11-26_Tables.R
\centering
\begin{tabular}{cccccccc}
\hline
 &  \multicolumn{2}{c}{Algorithm [KD]}   &       &   \\ \hline
$\nu$            & $\Psi_{12}$     & $\sigma_{11}$     & $\sigma_{12}$        \\
1                & 0               & 1.37 (1.19, 1.53) & -0.11 (-0.19, -0.02)  \\
3                & -0.5            & 1.33 (1.15, 1.50)  & -0.12 (-0.21, -0.04)   \\
3                & 0.5             & 1.38 (1.20, 1.53)  & -0.12 (-0.21, -0.03)  \\ \hline
    &              \multicolumn{2}{c}{Algorithm [P1]}         & &  \\ \hline                
$\nu$            & $\Psi_{12}$     & $\sigma_{11}$     & $\sigma_{12}$      \\   
1                & 1.14 (0.98, 1.29) & 0.39 (0.20, 0.56)  \\
3                & 1.14 (0.97, 1.31) & 0.34 (0.14, 0.51) \\
3                & 1.13 (0.98, 1.28) & 0.37 (0.19, 0.54) \\ \hline 
    &                \multicolumn{2}{c}{Algorithm [P2]} &&               \\ \hline
$\nu$            & $\Psi_{12}$     & $\sigma_{11}$     & $\sigma_{12}$       \\
1                & 0               & 1.13 (0.95, 1.30)  & 0.36 (0.20, 0.52)  \\
3                & -0.5             & 1.14 (0.99, 1.29) & 0.36 (0.17, 0.53)        \\
3                & 0.5              & 1.13 (0.96, 1.29) & 0.40 (0.21, 0.56)     \\ \hline
\end{tabular}
\caption{Comparison of Algorithms [KD], [P1], and [P2] under reference level 3, on $\Sigma$, the covariance of latent utilities under the trace constraint (equal to $C$), with rows corresponding to prior hyperparameters $(\nu-C,\Psi_{12})$ being $(1,0)$, $(3,-0.5)$, and $(3,0.5)$, respectively, where $\Psi$ is the scale matrix in the prior of $\widetilde{\Sigma}$ with $\Psi_{11}=\Psi_{22}=1$. Training and test datasets each with 5000 observations are generated under Setting 1 using reference level 3. }
\label{Table4}
\end{table}

\begin{table}[H]%Table 5 in BartGcomp_19-11-26_Tables.R
\centering
\begin{tabular}{cccccccc}
   &      & \multicolumn{6}{c}{Posterior Agreement Accuracy}          \\ \hline
   &      & \multicolumn{3}{c}{Train} & \multicolumn{3}{c}{Test} \\ \hline
$\nu - C$ & $\Psi_{12}$    & KD    & P1   & P2   & KD    & P1   & P2  \\ \hline
1  & 0    & 0.65    & 0.91   & 0.90   & 0.63    & 0.88   & 0.88  \\
3  & -0.5 & 0.70    & 0.91   & 0.90   & 0.67    & 0.88   & 0.88  \\
3  & 0.5  & 0.70    & 0.91   & 0.91   & 0.67    & 0.88   & 0.88  \\ \hline
   &      &         &        &        &         &        &       \\
   &      & \multicolumn{6}{c}{Posterior Mode Accuracy}          \\ \hline
   &      & \multicolumn{3}{c}{Train} & \multicolumn{3}{c}{Test} \\ \hline
$\nu - C$ & $\Psi_{12}$    & KD    & P1   & P2   & KD    & P1   & P2  \\ \hline
1  & 0    & 0.92    & 0.96   & 0.95   & 0.89    & 0.91   & 0.91  \\
3  & -0.5 & 0.93    & 0.95   & 0.95   & 0.90    & 0.91   & 0.91  \\
3  & 0.5  & 0.92    & 0.95   & 0.96   & 0.89    & 0.91   & 0.91  \\ \hline
\end{tabular}
	\caption{Accuracy comparison of Algorithms [KD], [P1], and [P2] under reference level 3. Training and test datasets each with 5000 observations are generated under Setting 2 with reference level 3. The prior of $\widetilde{\Sigma}$ is $\text{Inv-Wishart}(\nu,\Psi)$, where $\Psi_{11} = \Psi_{22} = 1$. Posterior predictive accuracy measured by \eqref{accmean} and \eqref{accmode} are reported under $(\nu-C,\Psi_{12})$ being $(1,0)$, $(3,-0.5)$, and $(3,0.5)$. }
	\label{Table5}
\end{table}

\begin{table}[H]%Table 6 in BartGcomp_19-11-26_Tables.R
\centering
\begin{tabular}{cccccccc}
\hline
      &              \multicolumn{2}{c}{Algorithm [KD]}          &  &  \\ \hline
$\nu$ & $\Psi_{12}$   & $\sigma_{11}$     & $\sigma_{12}$         \\
1     & 0             & 0.88 (0.64, 1.17) & -0.32 (-0.46, -0.18) \\
3     & -0.5         & 0.66 (0.47, 0.89) & -0.36 (-0.48, -0.23) \\
3     & 0.5           & 0.70 (0.50, 0.89) & -0.35 (-0.48, -0.21) \\ \hline
      &                  \multicolumn{2}{c}{Algorithm [P1]}   &&    \\ \hline
      1     & 0             & 0.79 (0.66, 0.92) & 0.82 (0.72, 0.89) \\
3     & -0.5        &  0.77 (0.64, 0.89) & 0.83 (0.75, 0.88) \\
3     & 0.5          & 0.80 (0.67, 0.93) & 0.80 (0.72, 0.87) \\ \hline
      &             \multicolumn{2}{c}{Algorithm [P2]}          &  &      \\ \hline
$\nu$ & $\Psi_{12}$  & $\sigma_{11}$     & $\sigma_{12}$                       \\
1     & 0            & 0.75 (0.63, 0.88) & 0.81 (0.73, 0.87)             \\
3     & -0.5        & 0.79 (0.61, 0.95) & 0.78 (0.68, 0.85)         \\
3     & 0.5          & 0.78 (0.64, 0.91) & 0.82 (0.74, 0.88)                \\ \hline
\end{tabular}
\caption{Comparison of Algorithms [KD], [P1], and [P2] under reference level 3, on $\Sigma$, the covariance of latent utilities under the trace constraint (equal to $C$), with rows corresponding to prior hyperparameters $(\nu-C,\Psi_{12})$ being $(1,0)$, $(3,-0.5)$, and $(3,0.5)$, respectively, where $\Psi$ is the scale matrix in the prior of $\widetilde{\Sigma}$ with $\Psi_{11}=\Psi_{22}=1$. Training and test datasets each with 5000 observations are generated under Setting 2 using reference level 3. }
\label{Table6}
\end{table}

\begin{table}[H] %BartGcomp_19-11-26_50data.R, on setting 9, 13, 17 in BartGcomp_19-11-26_Setting.R 
	\centering
	\begin{tabular}{ccccc}
	
		&      & \multicolumn{3}{c}{$E\{E[\sigma_{11}|D]\} \quad (S\{E[\sigma_{11}|D]\})$}          \\ \hline
		
		$\nu - C$ & $\Psi_{12}$    & KD    & P1   & P2   \\ \hline
		1  & 0    & 1.311 (0.032) & 1.035 (0.041) & 1.039 (0.039) \\
		3  & -0.5 & 1.325 (0.057) & 1.035 (0.036) & 1.034 (0.036)  \\
		3  & 0.5  & 1.296 (0.059) & 1.039 (0.042) & 1.038 (0.042) \\ \hline
		&      &     &       &        \\ 
		&      & \multicolumn{3}{c}{$E\{E[\sigma_{12}|D]\} \quad (S\{E[\sigma_{12}|D]\})$}                    \\ \hline
		$\nu - C$ & $\Psi_{12}$    & KD    & P1   & P2  \\ \hline
		1  & 0    & -0.108 (0.007) & 0.344 (0.053) & 0.354 (0.056)  \\
		3  & -0.5 & -0.118 (0.009) & 0.321 (0.057) & 0.348 (0.062) \\
		3  & 0.5  &  -0.122 (0.010) & 0.365 (0.054) & 0.359 (0.059) \\ \hline
	\end{tabular}
	\caption{Comparison of Algorithms [KD], [P1], and [P2] under reference level 3 on $\Sigma$ with 50 replications. Training and test datasets each with 5000 observations are generated under Setting 1 using reference level 3. $E[\cdot|D]$ indicates sample mean on one simulated data $D$. $E\{E[\cdot|D]\}$ and $S\{E[\cdot|D]\}$ are the mean and sd of $E[\cdot|D]$ across the 50 simulations of $D$. The prior of $\widetilde{\Sigma}$ is $\text{Inv-Wishart}(\nu,\Psi)$, where $\Psi_{11} = \Psi_{22} = 1$. Posterior predictive accuracy measured by \eqref{accmean} and \eqref{accmode} are reported under $(\nu-C,\Psi_{12})$ being $(1,0)$, $(3,-0.5)$, and $(3,0.5)$. }
	\label{Table7}
\end{table}

\begin{table*}
	\small
	\centering
	\begin{tabular}{cccccccc}
		&      & \multicolumn{6}{c}{Posterior Agreement Accuracy}                                                   \\ \hline
		&      & \multicolumn{3}{c}{Train}                     & \multicolumn{3}{c}{Test}                      \\ \hline
$\nu - C$ & $\Psi_{12}$    & KD    & P1   & P2   & KD    & P1   & P2  \\ \hline
		1  & 0    & 0.603 (0.003) & 0.900 (0.004) & 0.900 (0.004) & 0.580 (0.004) & 0.881 (0.003) & 0.882 (0.003) \\
		3  & -0.5 & 0.650 (0.004) & 0.900 (0.004) & 0.900 (0.004) & 0.618 (0.004) & 0.881 (0.003) & 0.882 (0.003) \\
		3  & 0.5  & 0.647 (0.003) & 0.900 (0.004) & 0.900 (0.004) & 0.617 (0.004) & 0.881 (0.003) & 0.882 (0.003) \\ \hline
		&      &               &               &               &               &               &               \\
		&      & \multicolumn{6}{c}{Posterior Mode Accuracy}                                                   \\ \hline
		&      & \multicolumn{3}{c}{Train}                     & \multicolumn{3}{c}{Test}                      \\ \hline
$\nu - C$ & $\Psi_{12}$    & KD    & P1   & P2   & KD    & P1   & P2  \\ \hline
		1  & 0    & 0.921 (0.005) & 0.946 (0.003) & 0.946 (0.003) & 0.872 (0.006) & 0.919 (0.004) & 0.919 (0.004) \\
		3  & -0.5 & 0.932 (0.004) & 0.946 (0.003) & 0.946 (0.003) & 0.872 (0.006) & 0.919 (0.003) & 0.919 (0.004) \\
		3  & 0.5  & 0.928 (0.004) & 0.946 (0.004) & 0.946 (0.003) & 0.873 (0.006) & 0.919 (0.004) & 0.919 (0.003) \\ \hline
	\end{tabular}
\caption{Accuracy comparison of Algorithms [KD], [P1], and [P2] under reference level 3 with 50 replications. Training and test datasets each with 5000 observations are generated under Setting 1 using reference level 3. Average (standard deviation) of accuracy across the 50 rounds are reported. The prior of $\widetilde{\Sigma}$ is $\text{Inv-Wishart}(\nu,\Psi)$, where $\Psi_{11} = \Psi_{22} = 1$. Posterior predictive accuracy measured by \eqref{accmean} and \eqref{accmode} are reported under $(\nu-C,\Psi_{12})$ being $(1,0)$, $(3,-0.5)$, and $(3,0.5)$. }
\label{Table8}
\end{table*}

\begin{table}[H] %BartGcomp_19-11-26_50data.R, on setting 9, 13, 17 in BartGcomp_19-11-26_Setting.R 
	\centering
	\begin{tabular}{ccccc}
		
		&      & \multicolumn{3}{c}{$E\{E[\sigma_{11}|D]\} \quad (S\{E[\sigma_{11}|D]\})$}          \\ \hline
		
		$\nu - C$ & $\Psi_{12}$    & KD    & P1   & P2   \\ \hline
1  & 0    & 0.848 (0.051) & 0.769 (0.046) & 0.770 (0.041)  \\
3  & -0.5 & 0.599 (0.059) & 0.783 (0.047) & 0.774 (0.041)  \\
3  & 0.5  & 0.691 (0.067) & 0.779 (0.036) & 0.758 (0.049)  \\ \hline
		&      &     &       &        \\ 
		&      & \multicolumn{3}{c}{$E\{E[\sigma_{12}|D]\} \quad (S\{E[\sigma_{12}|D]\})$}                    \\ \hline
		$\nu - C$ & $\Psi_{12}$    & KD    & P1   & P2  \\ \hline
1  & 0    & -0.321 (0.009) & 0.801 (0.029) & 0.797 (0.025) \\
3  & -0.5 & -0.349 (0.011) & 0.782 (0.028) & 0.791 (0.026) \\
3  & 0.5  & -0.354 (0.011) & 0.801 (0.026) & 0.802 (0.028) \\ \hline
	\end{tabular}
	\caption{Comparison of Algorithms [KD], [P1], and [P2] under reference level 3 on $\Sigma$ with 50 replications. Training and test datasets each with 5000 observations are generated under Setting 2 using reference level 3. $E[\cdot|D]$ indicates sample mean on one simulated data $D$. $E\{E[\cdot|D]\}$ and $S\{E[\cdot|D]\}$ are the mean and sd of $E[\cdot|D]$ across the 50 simulations of $D$. The prior of $\widetilde{\Sigma}$ is $\text{Inv-Wishart}(\nu,\Psi)$, where $\Psi_{11} = \Psi_{22} = 1$. Posterior predictive accuracy measured by \eqref{accmean} and \eqref{accmode} are reported under $(\nu-C,\Psi_{12})$ being $(1,0)$, $(3,-0.5)$, and $(3,0.5)$. }
	\label{Table9}
\end{table}

\begin{table*}
	\small
	\centering
	\begin{tabular}{cccccccc}
		&      & \multicolumn{6}{c}{Posterior Agreement Accuracy}                                                   \\ \hline
		&      & \multicolumn{3}{c}{Train}                     & \multicolumn{3}{c}{Test}                      \\ \hline
		$\nu - C$ & $\Psi_{12}$    & KD    & P1   & P2   & KD    & P1   & P2  \\ \hline
1  & 0    & 0.651 (0.003) & 0.905 (0.003) & 0.905 (0.003) & 0.632 (0.005) & 0.881 (0.003) & 0.882 (0.003) \\
3  & -0.5 & 0.701 (0.003) & 0.905 (0.003) & 0.905 (0.003) & 0.676 (0.005) & 0.882 (0.003) & 0.882 (0.003) \\
3  & 0.5  & 0.700 (0.003) & 0.906 (0.003) & 0.906 (0.004) & 0.679 (0.005) & 0.882 (0.003) & 0.882 (0.004) \\ \hline
		&      &               &               &               &               &               &               \\
		&      & \multicolumn{6}{c}{Posterior Mode Accuracy}                                                   \\ \hline
		&      & \multicolumn{3}{c}{Train}                     & \multicolumn{3}{c}{Test}                      \\ \hline
		$\nu - C$ & $\Psi_{12}$    & KD    & P1   & P2   & KD    & P1   & P2  \\ \hline
1  & 0    & 0.924 (0.004) & 0.953 (0.003) & 0.952 (0.003) & 0.896 (0.005) & 0.918 (0.004) & 0.918 (0.004) \\
3  & -0.5 & 0.930 (0.003) & 0.952 (0.003) & 0.952 (0.003) & 0.893 (0.006) & 0.918 (0.004) & 0.918 (0.004) \\
3  & 0.5  & 0.927 (0.003) & 0.952 (0.003) & 0.952 (0.003) & 0.895 (0.005) & 0.918 (0.004) & 0.918 (0.004) \\ \hline
	\end{tabular}
	\caption{Accuracy comparison of Algorithms [KD], [P1], and [P2] under reference level 3 with 50 replications. Training and test datasets each with 5000 observations are generated under Setting 2 using reference level 3. Average (standard deviation) of accuracy across the 50 rounds are reported. The prior of $\widetilde{\Sigma}$ is $\text{Inv-Wishart}(\nu,\Psi)$, where $\Psi_{11} = \Psi_{22} = 1$. Posterior predictive accuracy measured by \eqref{accmean} and \eqref{accmode} are reported under $(\nu-C,\Psi_{12})$ being $(1,0)$, $(3,-0.5)$, and $(3,0.5)$. }
	\label{Table10}
\end{table*}

\begin{table*}%Calculations.R from JHPCE
\centering
\begin{tabular}{ccccc}
                                  &                  & Disengaged (6497)       & Engaged (67462)      & Died (2697)     \\ \cline{3-5} 
Male                              &                  & 22.5                 & 34                   & 51.3               \\
Year of Enrolment                 & 2008             & 5.1                  & 9.7                  & 11.2               \\
                                  & 2009             & 8.3                  & 18.7                 & 17.1               \\
                                  & 2010             & 9.3                  & 17.3                 & 17.6               \\
                                  & 2011             & 9.2                  & 15.8                 & 17                 \\
                                  & 2012             & 17.9                 & 11.5                 & 14                 \\
                                  & 2013             & 18.5                 & 8.9                  & 11.3               \\
                                  & 2014             & 18.8                 & 9.0                  & 8.2                \\
                                  & 2015             & 12.8                 & 8.3                  & 3.3                \\
                                  & 2016             & 0.3                  & 0.8                  & 0.3                \\
Travel Time                       & \textless 30 min & 17.4                 & 24                   & 23.6               \\
                                  & 30 min - 1 h     & 19.4                 & 26.9                 & 29.4               \\
                                  & 1 h - 2 h        & 8.2                  & 14.6                 & 16.5               \\
                                  & \textgreater 2 h & 5.2                  & 7.7                  & 7.8                \\
                                  & Missing          & 49.9                 & 26.8                 & 22.6               \\
WHO Stage                         & 1                & 13.7                 & 4.7                  & 1.0                \\
                                  & 2                & 1.8                  & 2.0                  & 1.1                \\
                                  & 3                & 2.3                  & 2.2                  & 4.3                \\
                                  & 4                & 0.6                  & 0.3                  & 0.7                \\
                                  & Missing          & 81.7                 & 90.7                 & 92.9               \\
Married                           &                  & 57.2                 & 52.3                 & 49.7               \\
                                  & Missing          & 13.6                 & 8.3                  & 6.2                \\
On ART                            &                  & 39.9                 & 14.1                 & 14.1               \\
CD4 at/post baseline &                  & 64.8                 & 80.9                 & 74.7               \\
Post-baseline CD4 update      &                  & 6.6                  & 26                   & 7.6                \\
Most recent CD4                   &                  & 327 (144, 525)       & 279.77 (137, 462)    & 59 (18, 152)       \\
Age                               &                  & 29.91 (24.66, 36.51) & 35.56 (28.93, 43.65) & 37.97 (31.7, 45.7) \\
Height                            &                  & 163 (158, 169)       & 165 (159.1, 171)     & 167 (160, 173)     \\
                                  & Missing          & 24.6                 & 16.2                 & 17.7               \\
Weight                            &                  & 57.5 (51, 65)        & 56 (50, 63)          & 50 (44, 57)        \\
                                  & Missing          & 7.9                  & 3.9                  & 6.9                    \end{tabular}
\caption{Summary table of covariates stratified by outcome. The table reports ``median (25th percentile, 75th percentile)'' for continuous variables and percentage of true for binary variables or each level of categorical variables.
}
\label{Table11}
\end{table*}

\begin{table}[H]%Calculations.R from JHPCE
	\centering
\begin{tabular}{ccccccc}
	& \multicolumn{6}{c}{Posterior Agreement Accuracy}                                                                                             \\ \hline
	& \multicolumn{3}{c}{Train}                                          & \multicolumn{3}{c}{Test}                                           \\ \hline
	Ref Level            & KD                 & P1                 & P2                 & KD                 & P1                 & P2                 \\ \hline
	1                    & 0.67                 & 0.82                 & 0.82                 & 0.67                 & 0.81                 & 0.81                \\
	2                    & 0.55                 & 0.82                 & 0.82                 & 0.54                 & 0.81                 & 0.81                 \\
	3                    & 0.66                 & 0.82                 & 0.82                & 0.66                 & 0.81                 & 0.81                 \\ \hline
	\multicolumn{1}{l}{} & \multicolumn{1}{l}{} & \multicolumn{1}{l}{} & \multicolumn{1}{l}{} & \multicolumn{1}{l}{} & \multicolumn{1}{l}{} & \multicolumn{1}{l}{} \\
	& \multicolumn{6}{c}{Posterior Mode Accuracy}                                                                                             \\ \hline
	& \multicolumn{3}{c}{Train}                                          & \multicolumn{3}{c}{Test}                                           \\ \hline
	Ref Level            & KD                 & P1                 & P2                 & KD                 & P1                 & P2                 \\ \hline
	1                    & 0.88                 & 0.89                 & 0.89                 & 0.88                 & 0.89                 & 0.89                \\
	2                    & 0.85                 & 0.89                 & 0.89                  & 0.84                 & 0.89                 & 0.89                  \\
	3                    & 0.88                 & 0.89                 & 0.89                  & 0.88                 &0.89                 & 0.89                \\ \hline
\end{tabular}
	\caption{Accuracy comparison of Algorithms [KD], [P1], and [P2] on the AMPATH data. Posterior predictive accuracy measured by \eqref{accmean} and \eqref{accmode} are reported under reference levels 1, 2, and 3. The prior of $\widetilde{\Sigma}$ is $\text{Inv-Wishart}(3,I_3)$.}
	\label{Table12}
\end{table}

\section*{Figures}\label{fig}

\clearpage

\iffalse
\begin{figure*}
	\centering
\begin{tikzcd}
	W^{(t)} \arrow[dd, bend right] \arrow[ddd, bend right=49] & W^{(t+1)} \arrow[dd, bend left] \arrow[ddd, bend left=49] &  & W^{(t)} \arrow[ddd, bend right=49] \arrow[dd, bend right] & W^{(t+1)} \arrow[dd, bend left] \arrow[ddd, bend left=49] &  & W^{(t)} \arrow[ddd, bend right] \arrow[dd] & W^{(t+1)} \arrow[dd] \arrow[ddd, bend left] \\
\alpha^{(t)} \arrow[d,red] \arrow[dd, bend right]             & \alpha^{(t+1)} \arrow[red,d] \arrow[dd, bend left]            &  & \alpha^{(t)} \arrow[dd, bend right]                       & \alpha^{(t+1)} \arrow[dd, bend left]                      &  &                                            &                                             \\
	\mu^{(t)} \arrow[ruu] \arrow[d]                           & \mu^{(t+1)} \arrow[d]                                     &  & \mu^{(t)} \arrow[ruu] \arrow[d]                           & \mu^{(t+1)} \arrow[d]                                     &  & \mu^{(t)} \arrow[ruu] \arrow[d]            & \mu^{(t+1)} \arrow[d]                       \\
	\Sigma^{(t)} \arrow[ruuu] \arrow[ruu] \arrow[ru]          & \Sigma^{(t+1)}                                            &  & \Sigma^{(t)} \arrow[ruuu] \arrow[ruu] \arrow[ru]          & \Sigma^{(t+1)}                                            &  & \Sigma^{(t)} \arrow[ruuu] \arrow[ru]       & \Sigma^{(t+1)}                             
\end{tikzcd}
\caption{Above diagrams from left to right correspond to Algorithms [KD], [P1], and [P2], respectively.}\label{diagram}
\end{figure*}
\fi
\begin{figure*}
			\centering
			\includegraphics[scale=0.4]{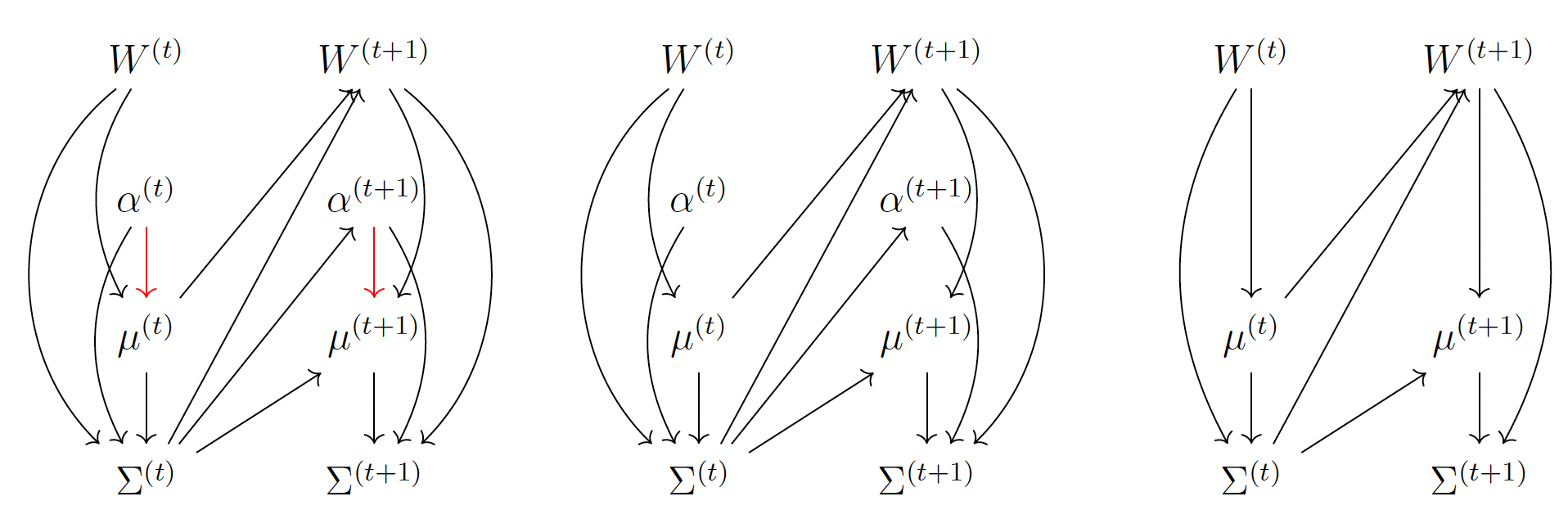}
			\caption{Above diagrams from left to right correspond to Algorithms [KD], [P1], and [P2], respectively.}\label{diagram}
\end{figure*}

\clearpage

%%%%%%%%%%%%%%%%%%%%%%%%%%%%%%%%%%%%%%%%%%%%%%
%% Single Appendix:                         %%
%%%%%%%%%%%%%%%%%%%%%%%%%%%%%%%%%%%%%%%%%%%%%%
\begin{appendix}
\section*{Appendix}
\renewcommand{\thesubsection}{\Alph{subsection}}

\subsection{Proof of Theorem 1}\label{ThmProof}

The lag-1 autocorrelation of $\mu$ is defined as 
$\text{corr}(\mu^{(t)}, \mu^{(t+1)})$ $= \frac{\text{cov}(\mu^{(t)}, \mu^{(t+1)})}{\sqrt{\text{var}(\mu^{(t)})\text{var}(\mu^{(t+1)})}},$
where $t$ indexes posterior draws. Under the condition that the chain has reached its stationary distribution\\
 $f(W,\mu,\Sigma,\alpha|S,X)$, where $S$ and $X$ are observed outcome and covariates, $\text{var}(\mu^{(t)})=\text{var}(\mu^{(t+1)})=\text{var}(\mu).$
Hence, we only need to look at the covariance for comparing the autocorrelation.

Consider two consecutive draws of $\mu$ from the Algorithm [P1]. We find that 
\begin{align*}
E(\mu^{(t)}\mu^{(t+1)}) =& E[E(\mu^{(t)}\mu^{(t+1)}|\Sigma^{(t)}, W^{(t+1)}, \alpha^{(t+1)})]\\
 =& E[E(\mu^{(t)}|\Sigma^{(t)}, W^{(t+1)}, \alpha^{(t+1)})]\\
& E[E(\mu^{(t+1)}|\Sigma^{(t)}, W^{(t+1)}, \alpha^{(t+1)})]\\
 = & E[E(\mu|\Sigma, W, \alpha)^2],
\end{align*}
where the first equality follows from the law of total expectation; the second and the third equalities follow from the fact that $\mu^{(t)}$ and $\mu^{(t+1)}$ are conditionally independent and identically distributed given $(\Sigma^{(t)}, W^{(t+1)}, \alpha^{(t+1)})$. This can be seen from the diagram for Algorithm [KD] in Figure \ref{diagram}; in particular, $\mu^{(t)}$ connects with $\mu^{(t+1)}$ only through $(\Sigma^{(t)}, W^{(t+1)}, \alpha^{(t+1)})$. As a result, the covariance is given by
\begin{align*}
\text{cov}(\mu^{(t)}, \mu^{(t+1)}) &= E(\mu^{(t)}\mu^{(t+1)}) - E(\mu^{(t)})E(\mu^{(t+1)})\\
&= E[E(\mu|\Sigma, W, \alpha)^2] - E[E(\mu|\Sigma, W, \alpha)]^2\\
&= \text{var}[E(\mu|\Sigma, W, \alpha)].
\end{align*}
Similarly, the covariance under Algorithms [P1] and [P2] is derived to be $\text{var}[E(\mu|\Sigma, W)]$. The key to this calculation is the fact that $\mu^{(t)}$ connects with $\mu^{(t+1)}$ only through $(\Sigma^{(t)}, W^{(t+1)})$ (see Figure \ref{diagram}). We now compare the two variances,
\begin{align*}
\text{var}[E(\mu|\Sigma, W, \alpha)] =& \text{var}\{E[E(\mu|\Sigma, W, \alpha)|\Sigma, W]\}\\
& + E\{\text{var}[E(\mu|\Sigma, W, \alpha)|\Sigma, W]\}\\
\ge & \text{var}\{E[E(\mu|\Sigma, W, \alpha)|\Sigma, W]\}\\
= &\text{var}[E(\mu|\Sigma, W)].
\end{align*}
The first equality comes from the law of total conditional variance and the last equality results from the law of total expectation. Therefore, we can conclude that the lag-1 autocorrelation of $\mu$, $\text{corr}(\mu^{(t)}, \mu^{(t+1)})$ is closer to zero in Algorithms [P1] and [P2] than in Algorithm [KD].

Recall from Figure \ref{diagram} that, $\Sigma^{(t)}$ connects with $\Sigma^{(t+1)}$ only through $(W^{(t+1)}, \alpha^{(t+1)}, \mu^{(t+1)})$ in Algorithms [KD] and [P1],  and through $(W^{(t+1)}, \mu^{(t+1)})$ in Algorithm [P2]. Hence, with a similar argument as above, we can show that the the lag-1 autocorrelation of $\Sigma$, $\text{corr}(\Sigma^{(t)}, \Sigma^{(t+1)})$, for Algorithm [P2] is no larger than in Algorithms [KD] and [P1].

\subsection{Equivalence in the Sampling Distribution between Algorithms [P1] and [P2]}\label{2eq3}

The purpose of this section is to show that Algorithms [P1] and [P2] have the same sampling distribution of $\Sigma$ when the sample size of the observed data, $N$, is sufficiently large and the scale matrix in the prior distribution of $\widetilde{\Sigma}$ is relatively small compared to the sample estimate of the covariance matrix for the latent variables. Note that Algorithms [P1] and [P2] draw $\theta$ and $W$ from the same conditional distributions. Hence, the conclusion here implies that the two procedures sample from the same joint distribution of $(\theta, W,\Sigma)$.

Given the prior on the unconstrained covariance matrix $\widetilde{\Sigma} \sim \text{Inv-Wishart}(\nu, \Psi)$, we can calculate in closed form the conditional posterior distributions of $\widetilde{\Sigma}$ and $\Sigma$ under Algorithms [P1] and [P2], respectively, where $\Sigma = \widetilde{\Sigma}\frac{C}{\text{trace}(\widetilde{\Sigma})}$. 

In Algorithm [P1], Step 1 samples the working parameter $\alpha^2_1$ from its conditional prior distribution, $$\alpha^2|\Sigma \sim \text{Inv-Gamma}(\nu C/2, \text{trace}(\Psi\Sigma^{-1})/2),$$
which is equivalent to $\text{trace}(\Psi \Sigma^{-1})/\chi^2_{\nu C}$\citep{imai_bayesian_2005} and has expectation $$E[\alpha^2_1|\Sigma] = \text{trace}(\Psi\Sigma^{-1})/(\nu C - 2).$$
Since inverse-Wishart is conditionally conjugate to the covariance matrix in a Gaussian model, the conditional posterior distribution of $\widetilde{\Sigma}$ under Algorithm [P1] is
$$\scriptstyle \widetilde{\Sigma}|\alpha^2_1, W, \mu \sim  \text{Inv-Wishart} (N+\nu, \Psi+\alpha^2_1\sum^N_{i=1} (W_i-\mu_i)(W_i-\mu_i)^T),$$ and this leads to the conditional posterior distribution of the corresponding restricted covariance matrix $\Sigma$,
\begin{align*} \textstyle
&f(\Sigma| \alpha^2_1, W, \mu)\propto |\Sigma|^{-(N+\nu+C+1)/2}\times \\
& \textstyle \text{trace}\bigg\{[\Psi+\alpha^2_1\sum^N_{i=1} (W_i-\mu_i)(W_i-\mu_i)^T]\Sigma^{-1}\bigg\}^{-\nu C/2}.
\end{align*} Similarly, the conditional posterior distributions under Algorithm [P2] for $\widetilde{\Sigma}$ and $\Sigma$, respectively, are
\begin{align*}
& \textstyle \widetilde{\Sigma}| W, \mu \sim  \textstyle \text{Inv-Wishart} (N+\nu, \Psi+\sum^N_{i=1} (W_i-\mu_i)(W_i-\mu_i)^T),\\
 &\textstyle f(\Sigma| W, \mu)\propto |\Sigma|^{-(N+\nu+C+1)/2}\times\\
  &\textstyle \text{trace}\bigg\{[\Psi+\sum^N_{i=1} (W_i-\mu_i)(W_i-\mu_i)^T]\Sigma^{-1}\bigg\}^{-\nu C/2}.
\end{align*} 
  
In order to compare the posterior conditional of $\Sigma$ under Algorithms [P1] and [P2], we look at the posterior mean and variance using a first-order Taylor series expansion. The posterior mean under Algorithm [P1] is
\begin{align*}
&E(\Sigma|\alpha^2_1, W, \mu) = E\bigg(\frac{\widetilde{\Sigma}}{\alpha^2}\bigg|\alpha^2_1, W, \mu\bigg)\approx \frac{E(\widetilde{\Sigma}|\alpha^2_1, W, \mu)}{E(\alpha^2|\alpha^2_1, W, \mu)}  \\
&= \frac{E(\widetilde{\Sigma}|\alpha^2_1, W, \mu)}{E[\text{trace}(\widetilde{\Sigma})|\alpha^2_1, W, \mu]}\times C\\
&= \frac{E(\widetilde{\Sigma}|\alpha^2_1, W, \mu)}{\text{trace}[E(\widetilde{\Sigma}|\alpha^2_1, W, \mu)]}\times C\\
&= \frac{\Psi+\alpha^2_1\sum^N_{i=1} (W_i-\mu_i)(W_i-\mu_i)^T}{\text{trace}[\Psi+\alpha^2_1\sum^N_{i=1} (W_i-\mu_i)(W_i-\mu_i)^T)]}\times C\\
&= \frac{\Psi+\alpha^2_1\sum^N_{i=1} (W_i-\mu_i)(W_i-\mu_i)^T}{\text{trace}(\Psi)+ \alpha^2_1\sum^N_{i=1}\sum^C_{j=1}(W_{ij}-\mu_{ij})^2}\times C
\end{align*}
Similarly, the posterior mean of $\Sigma$ under Algorithm [P2] is
$$E(\Sigma| W, \mu)=\frac{\Psi+\sum^N_{i=1} (W_i-\mu_i)(W_i-\mu_i)^T}{\text{trace}(\Psi)+ \sum^N_{i=1}\sum^C_{j=1}(W_{ij}-\mu_{ij})^2}\times C.$$

For the posterior variance of $\Sigma$, we simplify the notation by writing the posterior conditional distribution of $\widetilde{\Sigma}$ as $\text{Inv-Wishart}(\widetilde{\nu},\widetilde{\Psi})$, where $\widetilde{\nu} = N+\nu$ and the scale matrix $\widetilde{\Psi}$ is 
$$\left\{\begin{matrix}
\Psi+\alpha^2_1\sum^N_{i=1} (W_i-\mu_i)(W_i-\mu_i)^T & \text{ under Algorithm [P1]} \\ 
\Psi+\sum^N_{i=1} (W_i-\mu_i)(W_i-\mu_i)^T &  \text{ under Algorithm [P2]}
\end{matrix}\right.$$
Then, the posterior variance has the following form,
\begin{align*}
& \text{var}(\sigma_{ij}) = \text{var}(\frac{\widetilde{\sigma}_{ij}}{\alpha^2})\\
&\approx \frac{E(\widetilde{\sigma}_{ij})^2}{E(\alpha^2)^2}\times \bigg\{\frac{\text{var}(\widetilde{\sigma}_{ij})}{E(\widetilde{\sigma}_{ij})^2}-2\frac{\text{cov}(\widetilde{\sigma}_{ij},\alpha^2)}{E(\widetilde{\sigma}_{ij})E(\alpha^2)} + \frac{\text{var}(\alpha^2)}{E(\alpha^2)^2}\bigg\}  \\
\end{align*}where
\begin{align*}
E(\widetilde{\sigma}_{ij})&=\widetilde{\Psi}_{ij}/(\widetilde{\nu}-C-1)\\
E(\alpha^2)&= E(\widetilde{\sigma}_{11}+\ldots+\widetilde{\sigma}_{CC}) = \text{trace}(\widetilde{\Psi})/(\widetilde{\nu}-C-1)\\
\text{var}(\widetilde{\sigma}_{ij})&=\left\{\begin{matrix}
\frac{(\widetilde{\nu}-C+1)\widetilde{\Psi}_{ij}^2+ (\widetilde{\nu}-C-1)\widetilde{\Psi}_{ii}\widetilde{\Psi}_{jj}}{(\widetilde{\nu}-C)(\widetilde{\nu}-C-1)^2(\widetilde{\nu}-C-3)}& \text{ if } i\neq j\\ 
\frac{2\widetilde{\Psi}_{ii}^2}{(\widetilde{\nu}-C-1)^2(\widetilde{\nu}-C-3)}& \text{ if } i= j
\end{matrix}\right.\\
\text{var}(\alpha^2) &=  \frac{2}{(\widetilde{\nu}-C-1)^2(\widetilde{\nu}-C-3)} \sum^C_{i=1} \widetilde{\Psi}_{ii}^2\\
 \text{cov}(\widetilde{\sigma}_{ij},\alpha^2) &=  \sum^C_{k=1} \text{cov}(\widetilde{\sigma}_{ij},\widetilde{\sigma}_{kk}) \\ &= 2\frac{
	\widetilde{\Psi}_{ij}\widetilde{\Psi}_{kk} + (\widetilde{\nu}-C-1)\widetilde{\Psi}_{ik}\widetilde{\Psi}_{jk}
}{(\widetilde{\nu}-C)(\widetilde{\nu}-C-1)^2(\widetilde{\nu}-C-3)}.
\end{align*} 

Specifying the inverse-Wishart Prior for the covariance matrix is analogous to assuming a prior knowledge of $\nu$ Gaussian samples having covariance matrix $\Psi$. When the number of observations $N$ gets larger, the posterior concentrates more on the empirical covariance matrix. When $N$ is sufficiently large and $\Psi_{kj} << \alpha^2_1\sum^N_{i=1}\sum^C_{j=1}(W_{ik}-\mu_{ik})(W_{ij}-\mu_{ij})$ for all $k,j = 1,\ldots,C$, the posterior mean of $\Sigma$ under Algorithms [P1] and [P2] are approximately the same,
\begin{align*}
E(\Sigma|\alpha^2_1, W, \mu)&\approx \frac{\alpha^2_1\sum^N_{i=1} (W_i-\mu_i)(W_i-\mu_i)^T}{\alpha^2_1\sum^N_{i=1}\sum^C_{j=1}(W_{ij}-\mu_{ij})^2}\times C\\
&= \frac{\sum^N_{i=1} (W_i-\mu_i)(W_i-\mu_i)^T}{\sum^N_{i=1}\sum^C_{j=1}(W_{ij}-\mu_{ij})^2}\times C\\
&\approx E(\Sigma| W, \mu).
\end{align*}  
In the same way, we can show that $\text{var}(\Sigma|\alpha^2_1, W, \mu) \approx \text{var}(\Sigma| W, \mu)$.

\subsection{Outcome Distribution as a Function of Correlation}\label{Sbysig12}

This section discusses how the outcome distribution is connected to the correlation of latent utilities for $C=2$. We assume the outcome is determined as follows:
\begin{align*}
&W = (W_1,W_2)^T \sim \text{MVN}\bigg(\begin{pmatrix}
\mu_1\\ 
\mu_2
\end{pmatrix}, \left(\begin{matrix}
1& \rho\\
\rho & 1
\end{matrix}\right)\bigg),\\
&S = \left\{\begin{matrix}
1 & \text{if } W_1\ge \text{max}\{0,W_2\}\\
2 & \text{if } W_2\ge \text{max}\{0,W_1\}\\
3 & \text{if } W_1 <0 \text{ and } W_2 < 0.
\end{matrix}\right. 
\end{align*}

We start with the outcome probability at the reference level,
\vspace{-1em}
\begin{align*}
& P(S = 3) = P(W_1 <0 ,W_2 <0)\\
 &=\int^0_{-\infty}\int^0_{-\infty}\frac{1}{\sqrt{|2\pi \Sigma|}}\\
& \exp\bigg\{-\frac{1}{2}(w_1-\mu_1,w_2-\mu_2) \Sigma^{-1}\begin{pmatrix}
w_1-\mu_1\\ 
w_2-\mu_2
\end{pmatrix} \bigg\} dw_1dw_2\\
 &=\int^0_{-\infty}\int^0_{-\infty}\frac{1}{2\pi\sqrt{1-\rho^2}} \\
 &\exp\bigg\{ -\frac{1}{2} \frac{[w_1-\mu_1 - \rho (w_2-\mu_2)]^2}{1-\rho^2} -\frac{(w_2-\mu_2)^2}{2}\bigg\}  dw_1dw_2\\
 &= \int^{-\mu_2}_{-\infty}\int^0_{-\infty}\frac{1}{2\pi\sqrt{1-\rho^2}} \\
 &\exp\bigg\{ -\frac{1}{2} \frac{[w_1-\mu_1 - \rho \widetilde{w}_2]^2}{1-\rho^2} \bigg\} \exp\bigg\{-\frac{\widetilde{w}_2^2}{2}\bigg\}  dw_1d\widetilde{w}_2\\
&=\int^{-\mu_2}_{-\infty}\int^{\frac{-\mu_1-\rho \widetilde{w}_2}{\sqrt{1-\rho^2}}}_{-\infty}\frac{1}{2\pi} \exp\bigg\{ -\frac{\widetilde{w}^2_1}{2} \bigg\} \exp\bigg\{-\frac{\widetilde{w}^2_2}{2}\bigg\}  d\widetilde{w}_1 d\widetilde{w}_2\\
&= \int^{-\mu_2}_{-\infty} \frac{1}{2\pi} \tau\bigg(\frac{-\mu_1-\rho \widetilde{w}_2}{\sqrt{1-\rho^2}}\bigg) \exp\bigg\{-\frac{\widetilde{w}^2_2}{2}\bigg\} d\widetilde{w}_2,
\end{align*}
where the second equality comes from the inversion and determinant lemma of matrices, and $\tau(u) = \int^u_{-\infty} \exp\{-\frac{s^2}{2}\} ds$ in the last equality.

Next, we write $P(S = 3)$ as a function of $\rho$,\\
$f(\rho) = \int^{-\mu_2}_{-\infty} \frac{1}{2\pi} \tau(\frac{-\mu_1-\rho t}{\sqrt{1-\rho^2}}) \exp\{-\frac{t^2}{2}\} dt$. The corresponding derivative w.r.t $\rho$ has the form,
\begin{align*}
&\frac{d}{d\rho} f(\rho) \\
=& \int^{-\mu_2}_{-\infty} \frac{1}{2\pi\sqrt{1-\rho^2}} \\
&\exp\bigg\{-\frac{1}{2}\frac{(\mu_1 + \rho t)^2}{1-\rho^2} \bigg\} \exp\bigg\{-\frac{t^2}{2}\bigg\} \bigg(-\frac{t+\rho \mu_1}{1-\rho^2}\bigg)dt\\
 =& \int^{-\mu_2}_{-\infty} \frac{1}{2\pi\sqrt{1-\rho^2}} \\
&\exp\bigg\{-\frac{1}{2}\frac{(t+\rho\mu_1)^2 + \mu^2_1(1-\rho^2)}{1-\rho^2} \bigg\} \bigg(-\frac{t+\rho \mu_1}{1-\rho^2}\bigg)dt\\
 =& \int^{-\mu_2+\rho \mu_1}_{-\infty} \frac{1}{2\pi\sqrt{1-\rho^2}} \\
&\exp\bigg\{-\frac{1}{2}\frac{\widetilde{t}^2 }{1-\rho^2} \bigg\} \exp\bigg\{-\frac{1}{2}\mu^2_1\bigg\} \bigg(-\frac{\widetilde{t}}{1-\rho^2}\bigg)dt\\
 =& \frac{1}{2\pi\sqrt{1-\rho^2}} \exp\bigg\{-\frac{1}{2}\mu^2_1\bigg\}  \exp\bigg\{-\frac{1}{2}\frac{(-\mu_2+\rho \mu_1)^2}{1-\rho^2}\bigg\}\\
 \Rightarrow &\qquad  \frac{d}{d\rho} f(\rho) > 0, \qquad \rho\in (-1,1).
\end{align*}
As a result, for every possible combination of $(\mu_1,\mu_2)$, $P(S = 3)$ is always a strictly increasing function of $\rho$. Next, we show that how the non-reference-level outcome probabilities change w.r.t $\rho$ depends on $(\mu_1,\mu_2)$. For outcome level 1, we have 
\begin{align*}
P(S = 1) &= P(W_1 \ge W_2, W_1 \ge 0) = P(W_2-W_1 \le 0, -W_1 \le 0) \\
&= P(Z_1 \le 0, Z_2 \le 0)
\end{align*}
with $\begin{pmatrix}
Z_1\\ 
Z_2
\end{pmatrix} \sim N\bigg(\begin{pmatrix}
\mu_2 - \mu_1\\ 
- \mu_1
\end{pmatrix}, \left(\begin{matrix}
2(1-\rho) & 1-\rho\\
1-\rho & 1
\end{matrix}\right)\bigg)$. 

The outcome probability can be written as
\begin{align*}
&P(S = 1)  \\
 &= \textstyle\int^0_{-\infty}\int^0_{-\infty}\frac{1}{2\pi\sqrt{1-\rho^2}} \exp\bigg\{ -\frac{1}{2} \frac{[z_1-(\mu_2-\mu_1) - (1-\rho) (z_2+\mu_1)]^2}{1-\rho^2} \bigg\}\\
 & \exp\bigg\{-\frac{(z_2+\mu_1)^2}{2}\bigg\}  dw_1dw_2\\
 &= \int^{\mu_1}_{-\infty} \frac{1}{2\pi} \tau\bigg(\frac{-(\mu_2-\mu_1) - (1-\rho) t}{\sqrt{1-\rho^2}}\bigg)  \exp\bigg\{-\frac{t^2}{2}\bigg\}  dt
\end{align*}
Similar to the procedure for $f(\rho)$, we write $P(S = 1)$ as $g(\rho)$. The derivative w.r.t. $\rho$ is
\begin{align*}
 &\frac{d}{d\rho} g(\rho) =\\
 & \textstyle-\frac{1}{2} \exp\bigg\{-\frac{1}{1+\rho}\bigg[\frac{\mu_1 + \mu_2}{2}\bigg]^2 \bigg\}
- \frac{(\mu_2 - \mu_1)\sqrt{1+\rho}}{2(1-\rho)} \tau\bigg(\frac{\mu_1 + \mu_2}{2\sqrt{1+\rho}}\bigg)\\
\in&  \textstyle \bigg(-\frac{1}{2} \exp\bigg\{-\frac{1}{1+\rho}\bigg[\frac{\mu_1 + \mu_2}{2}\bigg]^2 \bigg\} \pm \frac{|\mu_1 - \mu_2|}{1-\rho}\sqrt{\frac{\pi(1+\rho)}{2}}\bigg).
\end{align*}
The last interval comes from $\tau(u) \in (0, \sqrt{2\pi})$ by definition. In fact, from the way $S$ depends on $(W_1, W_2)$ for the non-reference levels, we can easily see that the derivative of $P(S = 2)$ w.r.t $\rho$ falls into the same interval in the above derivation. Clearly, the center and width of the interval depend on $(\frac{\mu_1+\mu_2}{2},\rho)$ and $(|\mu_1-\mu_2|,\rho)$, respectively. So how $P(S = 1)$ and $P(S = 2)$ vary with $\rho$ are heavily influenced by the position of and the distance between the latent variables. 

The following plots display how the outcome distribution changes with $\rho$ under three different pairs, $(\mu_1,\mu_2)$. We can see that the reference level outcome probability $P(S = 3)$ increases with $\rho$ in all three settings, and the relative positions of $P(S=1)$ and $P(S=2)$ are closely related to the values of and the difference between $\mu_1$ and $\mu_2$.

\subsection{Algorithms' Pseudo Code}\label{Paper3Alg}

\subsubsection*{}
Using the notation in Section \ref{notation}, we provide details on the algorithms in Section \ref{algs}.

\begin{algorithm}[H]
\renewcommand{\thealgorithm}{}
	\small
	\caption{\textbf{[KD]}}
	\label{A1}
	%\begin{algorithmic}[1]
	\begin{algorithmic}
		\State \textbf{Step 0}: Initialize parameters $l=0, \alpha^{(0)},W^{(0)},\theta^{(0)},\Sigma^{(0)}$
		\While{$l < L$ do}
		\State \textbf{Step 1}: Update $(\widetilde{W}^{(l+1)}, (\alpha^{(l+1)}_1)^2)$ via $P(\widetilde{W},\alpha^2| \mu^{(l)},\Sigma^{(l)}, S)$ 
		\State \hspace{1em}(a) Draw $(\alpha^{(l+1)}_1)^2 \sim  \text{trace}[\Psi (\Sigma^{(l)})^{-1}]/\chi^2_{\nu C}$;
		\State \hspace{1em}(b) Draw $W^{(l+1)} = (W^{(l+1)}_{1},\ldots,W^{(l+1)}_{C})\sim MVN(\mu^{(l)}, \Sigma^{(l)})$ by
			\State \hspace{2.3em} \textbf{for} $i\in 1,\ldots,N $ \textbf{do}
			\State \hspace{3.6em} \textbf{for} $j \in 1,\ldots, C$ \textbf{do}
			\State \hspace{4.9em}  $W^{(l+1)}_{ij}|W^{(l+1)}_{i(-j)} \sim TN(m^{(l)}_{ij},(\tau^{(l)}_j)^2)$ \Comment{Appendix \ref{A11}}
			\State \hspace{4.9em} where $W^{(l+1)}_{i(-j)} = (W^{(l+1)}_{i1},\ldots,W^{(l+1)}_{i,j-1},W^{(l)}_{i,j+1},\ldots,W^{(l)}_{i,C})$
			\State \hspace{3.6em} \textbf{end for}
			\State \hspace{2.3em} \textbf{end for};
		\State \hspace{1em}(c) Set $\widetilde{W}^{(l+1)} = \alpha^{(l+1)}_1 W^{(l+1)}$.
		
		\State \textbf{Step 2}: Update $\widetilde{\theta}^{(l+1)}$ via $P(\widetilde{\theta}| \widetilde{W}^{(l+1)}, \alpha^{(l+1)}_1,\Sigma^{(l)})$ 
		\State \hspace{1em}(a) Gibbs sampling of binary trees: 
		\State \hspace{2.3em} \textbf{for} $b\in 1,\ldots,m $ \textbf{do}
			\State \hspace{3.6em} \textbf{for} $j \in 1,\ldots, C$ \textbf{do}
            \State \hspace{4.9em} Update $\widetilde{W}^{\dagger}_{jb}$ and draw  $\widetilde{\theta}^{(l+1)}_{jb}\sim P(\widetilde{\theta}_{jb}|\widetilde{W}^{\dagger}_{jb},(\alpha^{(l+1)}_1\tau^{(l)}_j)^2)$ \Comment{Appendix \ref{A12}}
			\State \hspace{3.6em} \textbf{end for}
			\State \hspace{2.3em} \textbf{end for};
		\State \hspace{1em}(b) Set $\widetilde{\mu}^{(l+1)}_{ij} = G_j(X_i;\widetilde{\theta}^{(l+1)}_j)$ and $\mu^{(l+1)}_{ij} = \widetilde{\mu}^{(l+1)}_{ij}/\alpha^{(l+1)}_1$.
		
		\State \textbf{Step 3}: Update $(\Sigma^{(l+1)}, (\alpha^{(l+1)}_3)^2)$ via $P(\Sigma, \alpha^2| \widetilde{W}^{(l+1)},\widetilde{\theta}^{(l+1)})$ 
		\State \hspace{1em}(a) Draw $\widetilde{\Sigma} \sim \text{Inv-Wishart}(n+\nu, \Psi+\sum^n_{i=1} \widetilde{\epsilon}_i\widetilde{\epsilon}^T_i)$,
		\State \hspace{2.3em}where $\widetilde{\epsilon}_i = (\widetilde{\epsilon}_{i1},\ldots, \widetilde{\epsilon}_{i,C})$ and $\widetilde{\epsilon}_{ij} = \widetilde{W}^{(l+1)}_{ij} - \widetilde{\mu}^{(l+1)}_{ij}$ for $j=1,\ldots,C$;
		
		\State \hspace{1em}(b) Setting $(\alpha^{(l+1)}_3)^2 = \text{trace}(\widetilde{\Sigma} / C)$;	
		\State \hspace{1em}(c) Re-scaling model parameters based on $\alpha^{(l+1)}_3$:
		\State \hspace{2.3em} $\Sigma^{(l+1)} = \widetilde{\Sigma} / (\alpha^{(l+1)}_3)^2$ and $W^{(l+1)} = \mu^{(l+1)} + \widetilde{\epsilon} / \alpha^{(l+1)}_3$;
		%\State \hspace{2.3em} \textbf{for} $i\in 1,\ldots,N $ \textbf{do}
			%\State \hspace{3.6em} \textbf{for} $j \in 1,\ldots, C$ \textbf{do}
			%\State \hspace{4.9em} $\mu^{(l+1)}_{ij} = \widetilde{\mu}_{ij} / \alpha^{(l+1)}$
			%\State \hspace{4.9em} $W^{(l+1)}_{ij} = \mu^{(l+1)}_{ij} + \widetilde{\epsilon}_{ij} / \alpha^{(l+1)}_3$
			%\State \hspace{3.6em} \textbf{end for}
			%\State \hspace{2.3em} \textbf{end for}
		\EndWhile
		\State \textbf{Step 4}: Prediction given new input \Comment{Appendix \ref{A13}}

	\end{algorithmic}
\end{algorithm}

\begin{algorithm}[H]
\renewcommand{\thealgorithm}{}
	\small
	\caption{\textbf{[P1]}}
	\label{A2}
	\begin{algorithmic}
		\State \textbf{Step 0}: Initialize parameters $l=0, \alpha^{(0)},W^{(0)},\theta^{(0)},\Sigma^{(0)}$
		\While{$l < L$ do}
		\State \textbf{Step 1}: Update $(\widetilde{W}^{(l+1)}, (\alpha^{(l+1)}_1)^2)$ via $P(\widetilde{W},\alpha^2| \mu^{(l)},\Sigma^{(l)}, S)$ 
		\State \hspace{1em}(a) Draw $(\alpha^{(l+1)}_1)^2 \sim  \text{trace}[\Psi (\Sigma^{(l)})^{-1}]/\chi^2_{\nu C}$;
		\State \hspace{1em}(b) Draw $W^{(l+1)} = (W^{(l+1)}_{1},\ldots,W^{(l+1)}_{C})\sim MVN(\mu^{(l)}, \Sigma^{(l)})$ by
			\State \hspace{2.3em} \textbf{for} $i\in 1,\ldots,N $ \textbf{do}
			\State \hspace{3.6em} \textbf{for} $j \in 1,\ldots, C$ \textbf{do}
			\State \hspace{4.9em}  $W^{(l+1)}_{ij}|W^{(l+1)}_{i(-j)} \sim TN(m^{(l)}_{ij},(\tau^{(l)}_j)^2)$ \Comment{Appendix \ref{A11}}
			\State \hspace{4.9em} where $W^{(l+1)}_{i(-j)} = (W^{(l+1)}_{i1},\ldots,W^{(l+1)}_{i,j-1},W^{(l)}_{i,j+1},\ldots,W^{(l)}_{i,C})$
			\State \hspace{3.6em} \textbf{end for}
			\State \hspace{2.3em} \textbf{end for};
		\State \hspace{1em}(c) Set $\widetilde{W}^{(l+1)} = \alpha^{(l+1)}_1 W^{(l+1)}$.
		
		\State \textbf{Step 2}: Update $\theta^{(l+1)}$ via $P(\theta| W^{(l+1)}, \Sigma^{(l)})$ 
		\State \hspace{1em}(a) Gibbs sampling of binary trees: 
		\State \hspace{2.3em} \textbf{for} $b\in 1,\ldots,m $ \textbf{do}
			\State \hspace{3.6em} \textbf{for} $j \in 1,\ldots, C$ \textbf{do}
            \State \hspace{4.9em} Update $W^{\dagger}_{jb}$ and draw  $\theta^{(l+1)}_{jb}\sim P(\theta_{jb}|W^{\dagger}_{jb},(\tau^{(l)}_j)^2)$ \Comment{Appendix \ref{A12}}
			\State \hspace{3.6em} \textbf{end for}
			\State \hspace{2.3em} \textbf{end for};
		\State \hspace{1em}(b) Set $\mu^{(l+1)}_{ij} = G_j(X_i;\theta^{(l+1)}_j)$.
		
		\State \textbf{Step 3}: Update $(\Sigma^{(l+1)}, (\alpha^{(l+1)}_3)^2)$ via $P(\Sigma, \alpha^2| \widetilde{W}^{(l+1)},\alpha^{(l+1)}_1,\theta^{(l+1)})$ 
		\State \hspace{1em}(a) Draw $\widetilde{\Sigma} \sim \text{Inv-Wishart}(n+\nu, \Psi+\sum^n_{i=1} \widetilde{\epsilon}_i\widetilde{\epsilon}^T_i)$,
		\State \hspace{2.3em}where $\widetilde{\epsilon}_i = (\widetilde{\epsilon}_{i1},\ldots, \widetilde{\epsilon}_{i,C})$ and $\widetilde{\epsilon}_{ij} = \widetilde{W}^{(l+1)}_{ij} - \alpha^{(l+1)}_1\mu^{(l+1)}_{ij}$ for $j=1,\ldots,C$;
		
		\State \hspace{1em}(b) Setting $(\alpha^{(l+1)}_3)^2 = \text{trace}(\widetilde{\Sigma} / C)$;	
		\State \hspace{1em}(c) Re-scaling model parameters based on $\alpha^{(l+1)}_3$:
		\State \hspace{2.3em} $\Sigma^{(l+1)} = \widetilde{\Sigma} / (\alpha^{(l+1)}_3)^2$ and $W^{(l+1)} = \mu^{(l+1)} + \widetilde{\epsilon} / \alpha^{(l+1)}_3$.

		\EndWhile
		\State \textbf{Step 4}: Prediction given new input \Comment{Appendix \ref{A13}}

	\end{algorithmic}
\end{algorithm}

\begin{algorithm}[H]
\renewcommand{\thealgorithm}{}
	\small
	\caption{\textbf{[P2]}}
	\label{A3}
	\begin{algorithmic}
		\State \textbf{Step 0}: Initialize parameters $l=0, \alpha^{(0)},W^{(0)},\theta^{(0)},\Sigma^{(0)}$
		\While{$l < L$ do}
		\State \textbf{Step 1}: Update $W^{(l+1)}$ via $P(W| \mu^{(l)},\Sigma^{(l)}, S)$ 
		\State \hspace{1em}(a) Draw $W^{(l+1)} = (W^{(l+1)}_{1},\ldots,W^{(l+1)}_{C})\sim MVN(\mu^{(l)}, \Sigma^{(l)})$ by
			\State \hspace{2.3em} \textbf{for} $i\in 1,\ldots,N $ \textbf{do}
			\State \hspace{3.6em} \textbf{for} $j \in 1,\ldots, C$ \textbf{do}
			\State \hspace{4.9em}  $W^{(l+1)}_{ij}|W^{(l+1)}_{i(-j)} \sim TN(m^{(l)}_{ij},(\tau^{(l)}_j)^2)$ \Comment{Appendix \ref{A11}}
			\State \hspace{4.9em} where $W^{(l+1)}_{i(-j)} = (W^{(l+1)}_{i1},\ldots,W^{(l+1)}_{i,j-1},W^{(l)}_{i,j+1},\ldots,W^{(l)}_{i,C})$
			\State \hspace{3.6em} \textbf{end for}
			\State \hspace{2.3em} \textbf{end for}.

		\State \textbf{Step 2}: Update $\theta^{(l+1)}$ via $P(\theta| W^{(l+1)}, \Sigma^{(l)})$ 
		\State \hspace{1em}(a) Gibbs sampling of binary trees: 
		\State \hspace{2.3em} \textbf{for} $b\in 1,\ldots,m $ \textbf{do}
			\State \hspace{3.6em} \textbf{for} $j \in 1,\ldots, C$ \textbf{do}
            \State \hspace{4.9em} Update $W^{\dagger}_{jb}$ and draw  $\theta^{(l+1)}_{jb}\sim P(\theta_{jb}|W^{\dagger}_{jb},(\tau^{(l)}_j)^2)$ \Comment{Appendix \ref{A12}}
			\State \hspace{3.6em} \textbf{end for}
			\State \hspace{2.3em} \textbf{end for};
		\State \hspace{1em}(b) Set $\mu^{(l+1)}_{ij} = G_j(X_i;\theta^{(l+1)}_j)$.
		
		\State \textbf{Step 3}: Update $(\Sigma^{(l+1)}, (\alpha^{(l+1)}_3)^2)$ via $P(\Sigma, \alpha^2| W^{(l+1)},\theta^{(l+1)})$ 
		\State \hspace{1em}(a) Draw $\widetilde{\Sigma} \sim \text{Inv-Wishart}(n+\nu, \Psi+\sum^n_{i=1} \epsilon_i\epsilon^T_i)$,
		\State \hspace{2.3em}where $\epsilon_i = (\epsilon_{i1},\ldots, \epsilon_{iC})$ and $\epsilon_{ij} = W^{(l+1)}_{ij} - \mu^{(l+1)}_{ij}$ for $j=1,\ldots,C$;
		
		\State \hspace{1em}(b) Setting $(\alpha^{(l+1)}_3)^2 = \text{trace}(\widetilde{\Sigma} / C)$;	
		\State \hspace{1em}(c) Re-scaling model parameters based on $\alpha^{(l+1)}_3$:
		\State \hspace{2.3em} $\Sigma^{(l+1)} = \widetilde{\Sigma} / (\alpha^{(l+1)}_3)^2$ and $W^{(l+1)} = \mu^{(l+1)} + \widetilde{\epsilon} / \alpha^{(l+1)}_3$.

		\EndWhile
		\State \textbf{Step 4}: Prediction given new input \Comment{Appendix \ref{A13}}

	\end{algorithmic}
\end{algorithm}

\subsubsection{Gibbs Sampling of the Latent Utilities}\label{A11}

Gibbs sampling of the latent utilities from univariate truncated normal distributions is described in the Section 3 of \citep{mcculloch_exact_1994}. In the pseudo-code, 
\vspace{-1em}
\begin{align*}
&m_{ij} = \mu_{ij} +\Sigma_{j(-j)}(\Sigma_{(-j)(-j)})^{-1}[W_{i(-j)}-\mu_{i(-j)})]\\
&\tau_j^2 = \Sigma_{jj}-\Sigma_{j(-j)}(\Sigma_{(-j)(-j)})^{-1}\Sigma_{(-j)j}
\end{align*}

for the $i=1,\ldots,N$, $j=1,\ldots,C$, where \\
$\mu_{ij}=\sum^m_{d = 1} g(X_i;\theta_{jd})$, $\Sigma_{jj}$ is the element at the $j$th row and $j$th column of $\Sigma$, $\Sigma_{(-j)(-j)}$ is the remaining of $\Sigma$ excluding its $j$th row and $j$th column, $\Sigma_{j(-j)}$ is the $j$th row of $\Sigma$ excluding its $j$th element $\Sigma_{jj}$, and  $\Sigma_{(-j)j}$ is similarly derived.

\subsubsection{Tree Sampling in MPBART}\label{A12}

Using Algorithms [P1] and [P2] as an example,we follow Section 3.2 of \citep{kindo_multinomial_2016} and provide the details on the conditional distributions used to update each individual tree. For simplicity, we exclude the subscript $i$. Given $W\sim MVN(\mu, \Sigma)$, we have $ W_{j}|(W_{(-j)},\mu,\Sigma) \sim N(m_{j},\tau^2_j)$ where $m_j$ and $\tau^2_j$ are defined in Appendix \ref{A11}. Based on the fact that $\mu_j = \sum^m_{d = 1} g(X;\theta_{jd})$, define
\begin{align*}
W^{\dagger}_{jb} =& W_{j} - \sum^{b-1}_{d = 1} g(X;\theta_{jd}) - \sum^{m}_{d' = b+1} g(X;\theta_{jd'})\\
& - \Sigma_{j(-j)}(\Sigma_{(-j)(-j)})^{-1}[W_{(-j)}-\mu_{(-j)})].
\end{align*}
Conditional on $(W_{(-j)},\mu,\Sigma)$,
\begin{align*}
&W^{\dagger}_{jb} - g(X;\theta_{jb}) = W_j -m_j \sim MVN(0,\tau^2_j) \\
 \Rightarrow &\quad W^{\dagger}_{jb}\sim N(g(X;\theta_{jb}), \tau^2_{j}).
\end{align*}
Consequently, the $b$th binary tree of the $j$th latent variable, $\theta^{(l+1)}_{jb}$, is updated to estimate the mean of    
\begin{align*}
W^{\dagger}_{jb} =& W^{(l+1)}_{j} - \sum^{b-1}_{d=1} g(X;\theta^{(l+1)}_{jd})- \sum^{m}_{d'= b+1} g(X;\theta^{(l)}_{jd'})\\
 &- \Sigma^{(l)}_{j(-j)}(\Sigma^{(l)}_{(-j)(-j)})^{-1}[W^{(l+1)}_{(-j)}- \mu_{(-j)}^{(l+1)}]
 \end{align*}	
where  $\mu_{(-j)}^{(l+1)}=\{G_k(X;\theta^{(l+1)}_k); k\neq j\}$ and $ G_k(X;\theta^{(l+1)}_k)= \sum^{b-1}_{d=1} g(X;\theta^{(l+1)}_{kd})+ \sum^{m}_{d'= b} g(X;\theta^{(l)}_{kd'})$.
Similarly, in Algorithm [KD], 
\begin{align*}
\widetilde{W}^{\dagger}_{jb} =& \widetilde{W}^{(l+1)}_{j} - \sum^{b-1}_{d=1} g(X\widetilde{\theta}^{(l+1)}_{jd})- \sum^{m}_{d'= b+1} g(X;\widetilde{\theta}^{(l)}_{jd'})\\
& - \Sigma^{(l)}_{j(-j)}(\Sigma^{(l)}_{(-j)(-j)})^{-1}[\widetilde{W}^{(l+1)}_{(-j)}- \widetilde{\mu}_{(-j)}^{(l+1)}]
\end{align*}		
where  $\widetilde{\mu}_{(-j)}^{(l+1)}=\{G_k(X;\widetilde{\theta}^{(l+1)}_k); k\neq j\}$ and $ G_k(X;\widetilde{\theta}^{(l+1)}_k)= \sum^{b-1}_{d=1} g(X;\widetilde{\theta}^{(l+1)}_{kd})+ \sum^{m}_{d'= b} g(X;\widetilde{\theta}^{(l)}_{kd'})$.

\subsubsection{Predictions from MPBART}\label{A13}

Given fitted model parameters from the $L$th round of posterior sampling, $(\theta^{(L)},\Sigma^{(L)})$, we obtain outcome prediction for a new input $x$ as follows:
\begin{align*}
	&	S(x)\\
		=&\left\{\begin{matrix}
		\text{reference level }0 &  \text{if }\max\{W_{1}(x),\ldots, W_C(x)\} < 0\\ 
		j &  \text{if }\max\{0,W_{1}(x),\ldots, W_C(x)\} = W_{j}(x)
		\end{matrix}\right.
		\end{align*}		
		by drawing
		$$W(x)\sim MVN(G(x;\theta^{(L)}), \Sigma^{(L)}).$$

\end{appendix}

%%%%%%%%%%%%%%%%%%%%%%%%%%%%%%%%%%%%%%%%%%%%%%
%% Multiple Appendixes:                     %%
%%%%%%%%%%%%%%%%%%%%%%%%%%%%%%%%%%%%%%%%%%%%%%
%\begin{appendix}
%\section{???}
%
%\section{???}
%
%\end{appendix}

%%%%%%%%%%%%%%%%%%%%%%%%%%%%%%%%%%%%%%%%%%%%%%
%% Support information, if any,             %%
%% should be provided in the                %%
%% Acknowledgements section.                %%
%%%%%%%%%%%%%%%%%%%%%%%%%%%%%%%%%%%%%%%%%%%%%%
\begin{acks}[Acknowledgments]
We would like to acknowledge support for this project
from the National Institutes of Health (NIH grants R01 AI 108441, R01 CA 183854, GM 112327,  AI 136664, K24 AI 134359). 
\end{acks}

\bibliographystyle{imsart-number}
 \bibliography{ref04-22}
 
%% or include bibliography directly:
% \begin{thebibliography}{}
% \bibitem{b1}
% \end{thebibliography}

\end{document}